\documentclass[11pt,reqno]{article}
\usepackage{amsmath,hyperref,amssymb, amsthm}
\usepackage{authblk}

\usepackage{graphicx,url}
\usepackage{color}
\usepackage{slashed}

\newcommand{\be}{\begin{eqnarray}}
\newcommand{\ee}{\end{eqnarray}}
\newcommand{\eeq}{\end{equation}}
\newcommand{\beq}{\begin{equation}}

\usepackage{simplewick}

\allowdisplaybreaks \numberwithin{equation}{section}

\DeclareSymbolFont{AMSa}{U}{msa}{m}{n}
\DeclareSymbolFont{AMSb}{U}{msb}{m}{n}
\DeclareMathSymbol{\fieldR}{\mathalpha}{AMSb}{"52}

\newcommand{\CH}{\mathcal{H}}

\newcommand{\CN}{\mathcal{N}}

\newcommand{\CQ}{\mathcal{Q}}

\DeclareMathOperator{\Tr}{Tr}
\DeclareMathOperator{\sTr}{sTr}

\newcommand{\NN}{\mathbb{N}}
\newcommand{\ZZ}{\mathbb{Z}}
\newcommand{\RR}{\mathbb{R}}
\newcommand{\CC}{\mathbb{C}}
\newcommand{\g}{\mathfrak{g}}

\def\t{\tau}

\def\beq{\begin{equation}}
\def\eeq{\end{equation}}
\def\bea{\begin{eqnarray}}
\def\eea{\end{eqnarray}}

\def\<{\langle}

\newcommand\nn{\nonumber}

\newtheorem{theorem}{Theorem}

\title{A Borcherds-Kac-Moody superalgebra with Conway symmetry}
\author[1, 2]{Sarah M. Harrison\thanks{sarharr@physics.mcgill.ca}}
\author[3]{Natalie M. Paquette\thanks{nataliep@caltech.edu}}
\author[4]{Roberto Volpato\thanks{volpato@pd.infn.it}}

{\small \affil[1]{\small Department of Mathematics and Statistics, McGill University, Montreal, QC, Canada}
\affil[2]{\small Department of Physics, McGill University, Montreal, QC, Canada}
\affil[3]{\small Walter Burke Institute for Theoretical Physics, California Institute of Technology,
Pasadena, CA 91125, USA}
\affil[4]{\small Dipartimento di Fisica e Astronomia `Galileo Galilei', Universit\`a di Padova \& INFN, sez. di Padova, Via Marzolo 8, 35131, Padova, Italy}}
\date{}

\begin{document}
\maketitle
\flushright{\small{CALT-TH-2018-005}}
\abstract{We construct  a Borcherds Kac-Moody (BKM) superalgebra on which the Conway group Co$_0$ acts faithfully. We show that the BKM algebra is generated by the BRST-closed states in a chiral superstring theory. We use this construction to produce denominator identities for the chiral partition functions of the Conway module $V^{s\natural}$, a supersymmetric $c=12$ chiral conformal field theory whose (twisted) partition functions enjoy moonshine properties and which has automorphism group isomorphic to Co$_0$. In particular, these functions satisfy a genus zero property analogous to that of monstrous moonshine.  Finally, we suggest how one may promote the denominators to spacetime BPS indices in type II string theory, which might thus furnish a physical explanation of the genus zero property of Conway moonshine.}

\newpage
\tableofcontents

\section{Introduction and motivation}\label{a:SVOA}

\textit{Motivation:} Our motivations for this work are twofold, each pertaining to an area of active interest in mathematical physics. The first is to provide a new, concrete example of an interesting infinite-dimensional algebra acting on BPS states in string theory. In this context, our algebra is a special type of Lie superalgebra known as a Borcherds Kac-Moody (BKM) algebra. We envision that this algebra will arise in a context close to the original proposal of Harvey and Moore \cite{HM1, HM2}: that BPS states form an algebra that in some situations is a BKM algebra.\footnote{Here, as in \cite{PPV, PPV2}, we expect to find a BKM algebra on the nose, such that its generators are not the BPS states themselves but rather are (conjecturally) in one-to-one correspondence with BPS states.} On the other hand, cohomological Hall algebras (COHAs) have been argued \cite{KS1, KS2} to be the correct framework to describe algebras of BPS states, and clarifying the relationship between BKM algebras and COHAs in explicit examples should be rewarding (see also \cite{KP} for a promising arena in which to relate these notions). It is clearly of interest to have more examples of algebras of BPS states---which moreover have deep connections to automorphic forms, motivic Donaldson-Thomas invariants, and vertex algebras---in the context of string compactifications on \textit{compact} Calabi-Yau manifolds or asymmetric (i.e. non-geometric) orbifolds of compact Calabi-Yau manifolds like those in \cite{PPV, PPV2} and this note. Our BKM is a concrete example in which we expect the relations among these topics can be explicitly studied, and we hope it forms an interesting new bridge among these subjects.

Our second motivation is to further clarify the relationship between moonshine and string theory. Moonshine in its most basic form is a surprising connection between sporadic groups and modular forms; it began with the observation that there is a connection between $J(\t)$, the unique modular function with respect to $SL(2,\ZZ)$ which has expansion\footnote{Throughout we take $ q= e^{2\pi i \t}$. }
$$ J(\t) \sim q^{-1} + O(q),$$
as $\t \to i \infty$, and the monster group, $\mathbb M$, the largest of the sporadic finite simple groups. This function has an expansion
$$
 J(\t) =\sum_{n=-1}^\infty c(n) q^n= q^{-1} + 196884 q + \ldots,
$$
  and satisfies the remarkable identity
\be\label{eq:denomJ}
p^{-1}\prod_{m>0, n\in \ZZ} (1-p^mq^n)^{c(mn)}= J(\sigma) - J(\t),
\ee
  where $p= e^{2\pi i \sigma}.$

The monstrous moonshine conjectures of Conway, Norton \cite{CN}, and Thompson \cite{Thompson1,Thompson2} state that there exists a $\ZZ$-graded vector space 
$$V^{\natural} = \bigoplus_{n=-1}^{\infty}V^{\natural}_n$$ 
such that each subspace $V^{\natural}_n$ furnishes a finite-dimensional representation of $\mathbb M$ and such that
$$J(\t) = \sum_{n=-1}^{\infty}\Tr_{V^{\natural}_n} q^n.$$
Furthermore, they conjecture that for each element $g \in \mathbb M$ the so-called \textit{McKay-Thompson series},
 $$T_g(\tau) := \sum_{n=-1}^{\infty}\text{Tr}_{V^{\natural}_n}(g) q^n,$$
is a Hauptmodul for a certain genus zero discrete subgroup $\Gamma_g < \text{SL}(2, \RR)$. 
We say that $\Gamma_g$ is \textit{genus zero} if the compactified quotient of the upper half-plane $\mathbb{H}$ by $\Gamma_g$ has the topology of a sphere and  that a modular function is a \textit{Hauptmodul} for $\Gamma_g$, when it establishes a biholomorphic map from $\overline{\mathbb{H}/\Gamma_g}$ to the Riemann sphere. 

A construction of $V^{\natural}$ was furnished by the work of Frenkel-Lepowsky-Meurmann (FLM) \cite{FLM} using the formalism of vertex operator algebra (VOA), or chiral conformal field theory (CFT) in physics terms.   The FLM monster module is a $c=24$ theory of chiral bosons with target space a $\ZZ_2$ orbifold  of the 24-dimensional torus $\RR^{24}/\Lambda$, where $\Lambda$ is the Leech lattice. (In fact, the construction in  \cite{FLM} represented the \emph{first} example of the procedure that is now known as orbifold of a  CFT.) The monstrous moonshine conjectures were proven by Borcherds \cite{BorcherdsMM} who made use of the FLM monster module $V^{\natural}$ and  the no-ghost theorem of bosonic string theory, and, moreover, who developed the formalism of BKM algebras to prove the genus zero property. Furthermore, the beautiful identity (\ref{eq:denomJ}) is interpreted as the (Borcherds)-Weyl-Kac denominator formula for this BKM algebra.

As we will review below, a proposal for a physical origin of the genus zero property was recently made in \cite{PPV, PPV2} by considering a certain two-dimensional compactification of heterotic string theory, where the monstrous BKM developed by Borcherds roughly appears as the an algebra of spacetime BPS states in this string theory. Thus the following natural questions arise: Is there a BKM algebra which plays a role in other instances of moonshine? And does it have an interpretation as an algebra of BPS states in string theory?

We answer the former question in the affirmative for the case of the so-called Conway moonshine, by constructing a (super)BKM algebra with Conway symmetry, and sketch a path to a positive answer for the second. As we review in \S \ref{sec:review}, Conway moonshine is a connection between the group $Co_0$ (``Conway zero"), the automorphism group of the Leech lattice, and a set of distinguished modular functions associated to elements $g \in Co_0$. Building on the work of \cite{FLM, Duncan}, in \cite{Duncan:2014eha} Duncan and Mack-Crane establish the existence of a ${1\over 2} \ZZ$-graded module dubbed $V^{s\natural}$ whose graded trace functions are normalized Hauptmoduln for genus zero groups. Thus $V^{s\natural}$ is a natural analogue of the monster module $V^\natural$. This module can be thought of as the Neveu-Schwarz sector of a chiral $\mathcal N=1$ superconformal field theory of central charge 12. Furthermore, there exists a canonically twisted module $V^{s\natural}_{tw}$, which can be thought of as the Ramond sector of this chiral theory, and whose trace functions also satisfy a genus zero property.

More specifically, if we define
$$ V^{s\natural}=\bigoplus_{n=-1}^{\infty}V^{s\natural}_{n/2},$$
and
$$V^{s\natural}_{tw}=\bigoplus_{n=0}^{\infty}V^{s\natural}_{tw,n},$$
we can associate the following set of graded (super)trace functions to each element $g\in Co_0$:\footnote{When $g$ is the identity element we suppress all subscripts $``g"$.}
\bea \label{eq:T1} T^s_g(\tau) &:=& \sum_{n=-1}^{\infty}\text{sTr}_{V^{s\natural}_{n/2}}(g) q^{n/2}=\sum_{n=-1}^\infty c^s_{g}(n/2) q^{n/2},\\\label{eq:T2}
\tilde T^s_g(\tau) &:=& \sum_{n=-1}^{\infty}\text{Tr}_{V^{s\natural}_{n/2}}(g) q^{n/2}=\sum_{n=-1}^\infty \tilde c^s_{g}(n/2) q^{n/2},\eea
and 
\bea \label{eq:T3}
T^s_{tw,g}(\tau) := \sum_{n=0}^{\infty}\text{sTr}_{V^{s\natural}_{tw,n}}(g) q^{n}=\sum_{n=0}^\infty c^s_{tw,g}(n) q^n,\\\label{eq:T4}
\tilde T^s_{tw,g}(\tau) := \sum_{n=0}^{\infty}\text{Tr}_{V^{s\natural}_{tw,n}}(g) q^{n}=\sum_{n=0}^\infty \tilde c^s_{tw,g}(n) q^n.
\eea
The  genus zero property of $V^{s\natural}$ was proven in Theorems 4.9 and 4.10 of \cite{Duncan:2014eha} and concerns the functions $T^s_g(\tau),T^s_{tw,g}(\tau)$. The explicit statements are the following:
\begin{enumerate}
\item  For all $g \in Co_0$,  the function $T^s_g(2\t)$ is a so-called ``normalized Hauptmodul" for a genus zero subgroup of $SL(2,\mathbb R)$; and
\item For all $g \in Co_0$, the function $T^s_{g,tw}(\t)$ is either 
\begin{enumerate}
\item constant with value $-\chi_g$ when $g$ has a fixed point in its action on the Leech lattice; or
\item a Hauptmodul for a genus zero subgroup of $SL(2,\mathbb R).$
\end{enumerate}
\end{enumerate}
By normalized Hauptmodul (with vanishing constant term), we mean a Hauptmodul which behaves as $T_g(q) \sim q^{-1}+ O(q)$ as $\t \to i\infty$. This property is the direct analogue of the genus zero property for the monstrous McKay-Thompson series. In fact, one finds that $T^s_{g,tw}(\t)=T_{g'}(\t)$ for some $g' \in \mathbb M$, and the corresponding genus zero group appearing is a  group $\Gamma_{g'}$ which appears in monstrous moonshine. Similarly, for many (though not all)  of the functions $T^s_g(2\t)$, the invariance group coincides with one of the genus zero groups appearing in monstrous moonshine.\footnote{We thank John Duncan for clarifications about this point.}

In this work we construct a super-BKM algebra with automorphism group $Co_0$ which is the natural analogue of the monstrous BKM algebra first constructed by Borcherds. We hope that the construction of this algebra may provide an independent route to understand (or, optimistically, provide an alternative proof of) the genus zero property of Conway moonshine, its corresponding role in string theory, and its connection to BPS algebras. In particular, we find (proven in \S \ref{sec:proof}) that the functions (\ref{eq:T1})-(\ref{eq:T4}) when $g$ is the identity satisfy a beautiful set of  formulas,
\bea\nonumber
&&p^{-1}\prod_{d=1}^\infty \prod_{r\in\ZZ} (1-p^{d}q^{r})^{c^s_{tw}({rd})} ={1\over \eta^{24}(\sigma)}-24-T^s_{tw}(\t)\\\nonumber
&&p^{-1}\prod_{d=1}^\infty \prod_{r\in \ZZ} (1-p^{d}q^{r})^{{(c^s_{tw}({rd})+\tilde c^s_{tw}({rd}))\over 2}}(1+p^{d}q^{r})^{{(c^s_{tw}({rd})-\tilde c^s_{tw}({rd}))\over 2}}=  {\eta^{24}(\sigma)\over \eta^{24}(2\sigma)}+24-\tilde T^{s}_{tw}(\t)\\\nonumber
&&p^{-1}\prod_{d=1}^\infty \prod_{r\in\ZZ} (1-p^{d}q^{\frac{r}{2}})^{C_{d,r}(\frac{rd}{2})} =  T^s( 2\sigma)-T^s(\t)\\\nonumber
&&p^{-1}\prod_{d=1}^\infty \prod_{r\in\ZZ} (1-(-1)^rp^{d}q^{\frac{r}{2}})^{C_{d,r}(\frac{rd}{2})} =  T^s(2\sigma)-\tilde T^s(\t),
\eea
where $C_{d,r}$ are defined in (\ref{eq:bigC}),\footnote{These $C_{d,r}$ can be expressed in terms of the coefficients $c^s, \tilde c^s$.} and where the first two lines correspond to the superdenominator and denominator of our algebra, respectively,\footnote{We conjecture that the last two lines represent the denominator and super-denominator of a different BKM superalgebra.} thus generalizing the famous Borcherds denominator formula (\ref{eq:denomJ}) for the $J$-function. Furthermore, we derive a twisted version of the first two formulas (see (\ref{Tw00}) and (\ref{Tw01})) for all elements $g \in Co_0$ by using the explicit action of the Conway group on our algebra. Finally, we conjecture corresponding twisted versions of the second two formulas (see (\ref{Tw10}) and (\ref{Tw11})) for all elements $g \in Co_0$, which should have an explanation in terms of an action of $Co_0$ on a different BKM superalgebra.

Finally, it is interesting to mention that the Conway moonshine module has also been of interest in the recent explorations of moonshine as it appears in string theory on K3, e.g. \cite{M5, DuncanMC, Equivariant, CHVZ, PVZ, HM3, KPV, CDR, Taormina:2017zlm}, and is therefore often closely linked with the Mathieu and umbral moonshine developments \cite{EOT, CDH1, CDH2}. Many aspects of the latter remain to be understood, including a uniform construction of explicit modules and their relationship to string theory/K3 surfaces, so we hope our explorations of the simpler, ``classical'' monstrous and Conway moonshines may be an important step towards building a unifying framework in which to address these open questions.

\textit{Background:} Not only does this  work rely heavily on the earlier works of Borcherds and Scheithauer \cite{BorcherdsMM, Sch1, Sch2}, but we are also motivated by the approach to monstrous moonshine developed in  \cite{PPV, PPV2}, which we briefly review below.

In \cite{PPV, PPV2}, the authors consider a heterotic string worldsheet theory where the internal or matter sector of the CFT is given by the $c=24$ Monster module of FLM on the left and the $c=12$ Conway module on the right. One may view this internal CFT as an asymmetric orbifold of a certain eight-torus. The theory is then compactified on an additional circle. From this basic theory, one can construct a family of theories by orbifolding---in fact, one can construct consistent orbifold theories for \textit{every} element of the Monster group (in contrast to orbifolds of the Monster module alone, which encounter failures of level-matching). The worldsheet analysis of these models provides a physical reinterpretation of the results of Borcherds and Carnahan \cite{BorcherdsMM, Carnahan, Carnahan2}: one tensors the matter CFT with an appropriate ghost system, applies the no-ghost theorem, and constructs a Lie algebra of physical states using Lie algebra homology (which has its physical origins in BRST cohomology). The generators and relations of this Lie algebra, manifested in the algebra's (twisted) denominator identities, imply that the McKay-Thompson series satisfy replicability identities that turn out to be equivalent to the Hauptmodul property \cite{Gannon}. Borcherds verified the last step using a brute-force computation to fix some low-lying coefficients of the modular functions in question. The analysis of \cite{PPV, PPV2} instead relates $\Gamma_g$ to the groups of string theory T-dualities, and provides a physical interpretation for the Hauptmodul property, which follows from a careful analysis of the behavior of the models upon decompactifying of the circle.
In the Conway case, we will follow a parallel worldsheet analysis by Scheithauer \cite{Sch1, Sch2, Sch5} which generalizes the work of Borcherds to SVOAs; indeed, the Conway module may be viewed as a  $\ZZ_2$-orbifold of the SVOA associated to the Fake Monster Lie algebra studied by Scheithauer. 

As a final point, we note that the denominator formulas of BKMs, including the Monster Lie algebra and the Conway Lie algebra of this note, are automorphic forms with respect to orthogonal groups of indefinite signature $O(m, n)$, e.g. $O(2, 2; \RR) \simeq SL(2, \RR)\times SL(2, \RR)$ \cite{borcherds1995automorphic, Borcherds:1996uda}\footnote{See also \cite{GN1, GN2} for examples of BKMs that are characterized by automorphic denominators.}. In \cite{PPV, PPV2}, these denominators naturally arise as BPS state counting functions in \textit{spacetime}, when one computes graded traces in the Fock space generated by an arbitrary number of free strings in the monstrous orbifold models. (In this note, we focus for simplicity on the worldsheet of a single string). Duality-invariant quantities in string theory, such as BPS state counting functions and certain couplings in the low energy effective action, are necessarily automorphic forms with respect to the duality group. In the context of K3 compactifications and orbifolds thereof, such functions are known to be essentially denominators of BKMs \cite{DVV, DJS} and may help clarify associated moonshine phenomena \cite{Cheng:2010pq,Persson:2013xpa,Persson:2015jka,PVZ}.

\textit{Summary:} The structure of the rest of the paper is as follows. In \S \ref{sec:review} we introduce the Conway module and describe its (twisted) partition functions in the Neveu-Schwarz and Ramond sectors. In \S \ref{sec:algebra} we  construct the Conway Lie algebra using the SVOA of Duncan, from the perspective of a  worldsheet of a chiral superstring theory. We introduce the matter and ghost sectors in turn, then discuss the BRST cohomology of physical states and the resulting Lie algebra. From this perspective we enumerate the multiplicity of roots in the algebra and derive  denominator  and superdenominator identities. Finally, we describe how our chiral construction can be completed to a type II string compactification (including, in particular, chiral and antichiral sectors and appropriate GSO projections). In \S \ref{sec:proof} we introduce equivariant Hecke operators and use this machinery to prove our (super)denominator identities; this proof is similar in approach to Borcherds' proof of the Koike-Norton-Zagier identity for the modular $J$-function, but requires some additional technology. In \S \ref{sec:simple} we use the denominator identity to deduce the simple roots of the Conway Lie algebra. Finally, we describe  twisted denominator formulas in \S \ref{sec:twisted}. These conjecturally correspond to BPS indices in a type II superstring compactification and would be essential in a physics proof of the genus zero property for Conway moonshine. We conclude with some open questions in \S \ref{sec:conc}. In four appendices we provide some technical details, respectively: a brief overview of Borcherds Kac-Moody algebras, some additional details on the construction of the algebra, computations regarding the zero-momentum limit of the BRST complex, and a proof of an identity used in the main text.

\section{The Conway modules $V^{f\natural}$ and $V^{s\natural}$}\label{sec:review}
In this section we review the description of certain distinguished super vertex operators algebras (SVOA), or chiral superconformal field theories in the physics language, with central charge 12 and symmetry group $Co_0$. We focus on two particular SVOAs: $V^{f\natural}$ and $V^{s\natural}$, which are discussed at length in \cite{Duncan} and \cite{Duncan:2014eha}, respectively. The former is related to what we call the ``internal" theory in our superstring construction in the next section, whereas the latter (or rather its canonically twisted module) is related to our BKM algebra after a GSO projection.

The vertex operator algebra $V^{f\natural}$ 
 can be described in terms of its bosonic subVOA, the vertex operator algebra (VOA) $V_{D_{12}}$, which is a lattice VOA based on the even lattice $D_{12}$. Equivalently, $V_{D_{12}}$ can be described as the basic representation of the Kac-Moody algebra $\hat{so}(24)_1$ at level $1$. The VOA $V_{D_{12}}$ has central charge $c=12$ with four irreducible modules, corresponding to the four cosets of $D_{12}^*/D_{12}$, which we denote by $A\cong V_{D_{12}}$ (the adjoint, or vacuum module, which is the VOA itself), $V$ (vector), $S$ (spinor) and $C$ (conjugate spinor).  
 
 The zero modes of the $so(24)_1$ currents generate a group of inner automorphisms isomorphic to $Spin(24)$, acting on $V_{D_{12}}\cong A$ and on its modules $S,V,C$. The center of $Spin(24)$ is $\ZZ_2 \times  \ZZ_2$. One $\ZZ_2$ acts by $-1$ on the spinor modules $S$ and $C$ and trivially on $A$ and $V$; it is the kernel of the covering map $Spin(24)\rightarrow SO(24)$. The generator of the second $\ZZ_2$ acts by $-1$ on $V$ and $C$ and trivially on $A$ and $S$; it is (a lift to $Spin(24)$ of) the center of $SO(24)$. 

Each of the three modules $V,S,C$ contains only fermionic states with $L_0$ eigenvalues (conformal weights) in $\frac{1}{2}+\ZZ$. In particular, the $V$ module contains $24$ fields of weight $1/2$, while the  $S$ and $C$ modules contain no fields of weight $1/2$.  One can extend the bosonic VOA $V_{D_{12}}$ by any one of these three ``fermionic'' modules to obtain an SVOA; the remaining two modules then form a ``canonically twisted module'' for the SVOA. (A remark: by SVOA we mean a vertex algebra including both bosons and fermions and with a stress-energy tensor; we do not include in the definition the choice of a supercurrent generating the $\cal N=1$ superconformal algebra. We will say explicitly when we make a choice for such a supercurrent).  In particular:

\begin{itemize}
	\item If we extend the bosonic VOA $V_{D_{12}}$ by the spinor module $S$, then we obtain the SVOA $V^{f \natural}$, introduced in \cite{FLM0} and studied in \cite{Duncan},
	\be V^{f \natural}\cong A\oplus S\ . \ee This is the unique (up to isomorphism) holomorphic self-dual SVOA of $c=12$ with no fields of weight $1/2$ \cite{Duncan}. It can also be described as the lattice SVOA based on the odd unimodular lattice $D_{12}^+$ containing $D_{12}$ and its $S$ coset. The SVOA $V^{f\natural}$ contains an $\cal N=1$ superVirasoro algebra, generated by a supercurrent $G(z)$, which is a suitable field of weight $3/2$ in the $S$ module of $V_{D_{12}}$. The field $G(z)$ of spin $3/2$ generating the $N=1$ superVirasoro algebra is unique up to $Spin(24)$ transformations; we choose one of them once and for all.  The  subgroup of $Spin(24)$ fixing $G(z)$ is isomorphic to the Conway group $Co_0$. More precisely, the group acting non-trivially on the SVOA $V^{f \natural}$  is $Co_1\cong Co_0/ \ZZ_2$, the quotient of $Co_0$ by its centre $\ZZ_2$. Indeed, the generator of the central $\ZZ_2$ in $Co_0\subset Spin(24)$ coincides with the generator of the second $\ZZ_2$ in the center of $Spin(24)$, the one acting by $-1$ on $V$ and $C$, i.e. the center of $SO(24)$ (lifted to $Spin(24)$).    Thus, the central element of $Co_0$ acts trivially on the cosets $A$ and $S$ comprising $V^{f \natural}$, and only the quotient $Co_1\cong Co_0/\ZZ_2$ acts faithfully. The SVOA $V^{f \natural}$ has a canonically twisted module $V^{f \natural}$ (a twisted sector with respect to the generator of the  central $\ZZ_2\subset Spin(24)$ acting by $-1$ on $S$ and by the identity on $A$) given by  
	\be V^{f\natural}_{tw}\cong V\oplus C\ .\ee On this twisted module the central element of $Co_0$ acts non-trivially as $-{\rm id}$. To summarize, $Co_0$ acts faithfully on the twisted module $V^{f\natural}_{tw}$, but only $Co_1$ acts faithfully on the $\cal N=1$ SVOA $V^{f\natural}$.
	\item If we extend the bosonic VOA $V_{D_{12}}$ by the conjugate spinor module $C$, we get the SVOA  
	\be V^{s\natural}\cong A\oplus C,\ee introduced in \cite{Duncan:2014eha}.
	As a SVOA it is isomorphic to $V^{f\natural}$, and in particular it contains no spin $1/2$ fields. It has a canonically twisted module $V^{s\natural}_{tw}$ (twisted with respect to the $\ZZ_2$ symmetry acting by $-1$ on $C$) given by 
	\be V^{s\natural}_{tw}\cong
	V\oplus S\ .\ee  
	The main difference with $V^{f\natural}$ is that the distinguished spin $3/2$ field $G(z)$ fixed by $Co_0$ now is not contained in the SVOA $V^{s\natural}$, but in its twisted module $V^{s\natural}_{tw}$. Thus, $G(z)$ cannot be identified with the generator of a $N=1$ superVirasoro algebra as in $V^{f \natural}$, but it can be interpreted here as a ``spectral flow generator'', an intertwining operator mapping the SVOA $V^{s\natural}$ to its canonically twisted module $V^{s\natural}_{tw}$. The group $Co_0$ fixing $G(z)$ acts faithfully both on the SVOA $V^{s\natural}$ and on its twisted module $V^{s\natural}_{tw}$: the central element of $Co_0$ acts non-trivially on the $C$ sector of the SVOA and on the $V$ sector of its twisted module.
	\item Finally, if we extend the bosonic VOA by the vector module $V$, we obtain the SVOA 
	\be F_{24}\cong A\oplus V\ ,\ee
	generated by $24$ free fermions. It can also be described as the lattice SVOA based on the odd unimodular lattice $\ZZ^{12}$. The two SVOAs $V^{f\natural}$ and $V^{s\natural}$ can  also be described as suitable $\ZZ_2$-orbifolds of $F_{24}$.
\end{itemize}

Having in mind the superstring construction in the next section, we will denote the Hilbert spaces of the four modules $A,V,S,C$ of $V_{D_{12}}$ as ${\rm NS}+$, ${\rm NS}-$, ${\rm R}+$ and ${\rm R}-$, respectively, such that
\be V_{\rm NS+}=A\qquad V_{\rm NS-}=S\qquad V_{\rm R+}=C\qquad V_{\rm R-}=V\ .
\ee In our notation, we define $V_{\rm NS}:=V_{\rm NS+}\oplus V_{\rm NS-} =V^{f\natural}$ and $V_{\rm R}:=V_{\rm R+}\oplus V_{\rm R-}=V^{f\natural}_{tw}$. We define the fermion number on $V_{\rm NS}$ and $V_{\rm R}$ as the symmetry in the central $\ZZ_2\times \ZZ_2\subset Spin(24)$ acting by $-id$ on the cosets $S$ and $V$ and trivially on $A$ and $C$.
The characters of each of these modules, with and without fermion number insertions, are given by
\bea\label{eq:partfns}
f_{\rm NS}(\tau)&:=& \Tr_{\rm NS}(q^{L_0 - 1/2}) = \frac{\eta^{48}(\t)}{\eta^{24}(\t/2)\eta^{24}(2\t)}-24\\
f_{\tilde{\rm NS}}(\tau)&:=& \sTr_{\rm NS}( q^{L_0 - 1/2}) =  {\eta^{24}(\t/2)\over \eta^{24}(\t)}+24\\
f_{\rm R}(\tau)&:=& \Tr_{\rm R}(q^{L_0 - 1/2})=  2^{12}{\eta^{24}(2\t)\over \eta^{24}(\t)}+24
\\
f_{\tilde{\rm R}}(\tau)&:=& \sTr_{\rm R}(q^{L_0 - 1/2})=-24\ ,
\eea
where by $\sTr$ we denote the supertrace; i.e. the trace with the insertion of the fermion number operator $(-1)^F.$ We also make use of the Dedekind eta function, defined as
\be
\eta(\tau):=q^{1/24}\prod_{n=1}^\infty (1-q^n).
\ee
The first several terms in the $q$-expansion of each of these functions are
\bea\label{eq:qexp}
f_{\rm NS}(\tau)&=& q^{-1/2} + 0 + 276 q^{1/2} + 2048 q + 11202 q^{3/2} + \ldots \\
f_{\tilde{\rm NS}}(\tau)&=& q^{-1/2} + 0 + 276 q^{1/2} - 2048 q + 11202 q^{3/2} + \ldots \\
f_{\rm R}(\tau)&=&  24 + 4096 q + 98304 q^2 + 1228800 q^3 
+ \ldots\\
f_{\tilde{\rm R}}(\tau)&=&-24\ .
\eea
For convenience, we also introduce the functions $F_{\rm NS+}$, $F_{\rm NS-}$, $F_{\rm R+}$, $F_{\rm R-}$ which are the projections in each sector onto states invariant or anti-invariant under $(-1)^F$:
\begin{align} &F_{\rm NS\pm}(\tau)=\Tr_{\rm NS}\left(\frac{1\pm (-1)^F}{2}q^{L_0 - 1/2}\right)=\frac{f_{\rm NS}\pm f_{\tilde{\rm NS}}}{2}\ ,\\
&F_{\rm R\pm}(\tau)=\Tr_{\rm R}\left(\frac{1\pm (-1)^F}{2}q^{L_0 - 1/2}\right)=\frac{f_{\rm R}\pm f_{\tilde{\rm R}}}{2}\ ;
\end{align} and we denote by $C_{\rm NS\pm}(r)$ and $C_{\rm R\pm}(n)$ the corresponding Fourier coefficients
\begin{align}
&F_{\rm NS\pm} (\tau)=\sum_{r\in \frac{1}{2}\ZZ} C_{\rm NS\pm}(r) q^r\ ,\\
&F_{\rm NS\pm} (\tau)=\sum_{n\in \ZZ} C_{\rm R\pm}(n) q^n\ ,
\end{align}
Note that
\be F_{\rm NS-}=F_{\rm R+}=F_{\rm R-}-24= 2048 q + 49152 q^2 + \ldots \ .
\ee

We will also introduce traces in the modules $V^{s\natural}$ and $V^{s\natural}_{tw}$, following \cite{Duncan:2014eha}. Anticipating notation we will use in \S \ref{sec:proof}, we define the following traces, which can be written in terms of the functions introduced above as
\bea\label{eq:Vsnat1}
f(1,1;\t)&=& -\Tr_{V^{s\natural}} q^{L_0-c/24}=-f_{\rm NS}(\tau)\\\label{eq:Vsnat2}
f(1,0;\t)&=& \sTr_{V^{s\natural}} q^{L_0-c/24}=f_{\tilde{\rm NS}}(\tau)\\\label{eq:Vsnat3}
f(0,1;\t)&=& \Tr_{V^{s\natural}_{tw}}  q^{L_0-c/24}=f_{\rm R}(\tau)\\\label{eq:Vsnat4}
f(0,0;\t)&=& \sTr_{V^{s\natural}_{tw}}  q^{L_0-c/24}=f_{\tilde{\rm R}}(\tau)
\eea

\section{Construction of the algebra}\label{sec:algebra}

The starting point of our construction is a chiral superconformal field theory (SCFT), or super vertex algebra (SVA)\footnote{A (super-)vertex algebra satisfies weaker conditions than a (super-)vertex operator algebra. In particular, the (super-)VA we will consider contains a stress-energy tensor, but the $L_0$ eigenspaces will not necessarily be finite dimensional, as required for a (super-)VOA.}, with a matter part of central charge $15$, and a standard ghost ($bc$-system) and superghost ($\beta\gamma$-system) part. Our construction may be viewed as the ``chiral half'' of the theory on the worldsheet of a particular type II string compactification, as we will discuss in \S \ref{sec:strings}. We impose a GSO projection on this chiral non-unitary CFT, and then consider a physical state condition, which means we study the cohomology of a nilpotent BRST charge $Q$. Following Lian and Zuckerman \cite{LZ}, we show that the physical states (i.e. the elements in the BRST cohomology) generate an infinite-dimensional generalized Kac-Moody superalgebra with $Co_0$ symmetry. 

\subsection{Matter sector} We will first describe the matter conformal field theory of $c=15$. The Neveu-Schwarz (NS) sector $\CH_{\rm NS}^m$ consists of the tensor product 
\be \CH_{\rm NS}^m=V_{\rm NS}\otimes V_{1,1}\ ,\ee of two vertex algebras of central charges $12$ and $3$, respectively. The former is the SVOA $V_{\rm NS}\cong V^{f\natural}$ introduced in the previous section.\footnote{Note that in this section we will denote the supercurrent and stress-energy tensor in $V^{f\natural}$ by $G'(z)$ and $T'(z)$, respectively.} . 

One can think of $V_{1,1}$ as describing the chiral half of a non-linear sigma model to a $1+1$ dimensional space-time compactified on $\RR^{2}/\Gamma^{1, 1}$. $\Gamma^{1, 1}$ is the unique even unimodular lattice of signature $(1, 1)$ and $V_{1, 1}$ is a non-unitary self-dual SVA of central charge $3$. This SVA is generated by the currents $\partial X^\mu=\sum_{n\in \ZZ}\alpha^\mu_nz^{-n-1}$, $\mu=0,1$, the fields $e^{ikX}$, $k\in \Gamma^{1,1}$, of weight $k^2/2$ (which form the bosonic subalgebra) and the fermions $\psi^\mu=\sum_r \psi^\mu_rz^{-r-\frac{1}{2}}$, $\mu=0,1$, of weight $1/2$, the superpartners of the supercurrents, with OPE
\be \psi^\mu(z)\psi^\nu(0)\sim \frac{\eta^{\mu\nu}}{z}\ ,\qquad \mu,\nu=0,1\ ,
\ee where $\eta^{\mu\nu}=\left(\begin{smallmatrix} -1 & 0\\ 0 & 1\end{smallmatrix} \right)$ is the spacetime metric. 
 The tensor product $V^{f\natural}\otimes V_{1,1}$ contains an $\CN=1$ superVirasoro algebra generated by
\be G_{r}:=\sum_{n\in \ZZ} \alpha_n^\mu \psi_{\mu (r-n)}+G'_r
\ee
\be L_n:=\frac{1}{2}\sum_{m\in \ZZ} :\alpha^\mu_m\alpha_{\mu(n-m)}:+\frac{1}{4}\sum_{r} (2r-n) :\psi^\mu_r\psi_{\mu(n-r)}:+L_n'
\ee where $G_r'$ and $L_n'$ are the superVirasoro generators of $V^{f\natural}$. We denote by $(-1)^F$ the standard fermion number of the SVA $V^{f\natural}\otimes V_{1,1}$, which acts in the NS sector by multiplication by $-1$ the fields with half-integral spin. This $\ZZ/2\ZZ$ symmetry is the product of the fermion number of the internal SVOA $V^{f\natural}$ and of the spacetime\footnote{This terminology comes from the very similar compactification in \cite{PPV, PPV2}, where one direction of the  transverse spacetime was compactified on a (large) $S^1$ and the Monster VOA $V^{\natural}$ furnished the chiral half of the internal CFT.} VA $V_{1,1}$.

The Ramond sector $\CH_{\rm R}^m$ of the theory consists of the tensor product of $V_{tw}^{f\natural}$, the unique canonically twisted module for $V^{f\natural}$, and the Ramond sector of $V_{1,1}$ (itself a canonically twisted module in the VA parlance). The fermion number $(-1)^F$ induces a $\ZZ_2$ symmetry on the Ramond sector.   We can therefore split the matter CFT into four sectors, namely $\CH_{\rm NS+}^m,\CH_{\rm NS-}^m,\CH_{\rm R+}^m,\CH_{\rm R-}^m$ that are eigenspaces of $(-1)^F$. It is often useful to consider the bosonizations of the  space-time fermions $\psi^\mu$
\be \psi^{\pm}\sim e^{\pm\lambda}\ ,\qquad :\psi^+\psi^-:\sim \partial\lambda\ ,
\ee where $\psi^{\pm}=\psi^0\pm\psi^1$, $\lambda$ is a chiral free boson which appears either derivated $\partial^n\lambda$, $n>0$ or in the exponential $e^{\frac{m}{2}\lambda}$, where $m\in \ZZ$. In particular, the ${\rm NS}$ and ${\rm R}$ sectors are characterized by $m$ being, respectively, even or odd. The matter fermion number $(-1)^F$ is obtained by multiplying the fermion number of the internal CFT $V^{f\natural}$ (${\rm NS}$ sector) or $V^{f\natural}_{tw}$ (${\rm R}$ sector) by $(-1)^{m/2}$ (${\rm NS}$ sector) or  $(-1)^{(m-1)/2}$ (${\rm R}$ sector).

\subsection{Ghosts and superghosts} Let us now consider the ghost and superghost sector $\CH^{gh}$. This is just the standard superstring theory construction: we introduce a $bc$ (ghost) system, the vertex algebra of two anticommuting fields $b(z)=\sum_n b_nz^{-n-2}$ and $c(z)=\sum_n c_nz^{-n+1}$ of conformal weights $2$ and $-1$ with OPE
\be b(z)c(w)\sim  \frac{1}{z-w}\qquad c(z)b(w)\sim  \frac{1}{z-w}\qquad b(z)b(w)\sim 0\qquad c(z)c(w)\sim 0\ ,
\ee and  stress tensor
\be T_{bc}(z)=-:(\partial b) c:(z)-2 :b\partial c:(z)
\ee  and the $\beta\gamma$ (superghost) system, the vertex algebra of two commuting fields $\beta(z)=\sum_r \beta_rz^{-r-\frac{3}{2}}$, $\gamma(z)=\sum_r \gamma_rz^{-r+\frac{1}{2}}$ of conformal weights $3/2$ and $-1/2$, with OPE
\be \beta(z)\gamma(w)\sim -\frac{1}{z-w}\qquad \gamma(z)\beta(w)\sim \frac{1}{z-w}\qquad \beta(z)\beta(w)\sim 0\qquad \gamma(z)\gamma(w)\sim 0
\ee and stress tensor
\be T_{\beta\gamma}=-\frac{1}{2}:(\partial\beta)\gamma:(z)-\frac{3}{2}:\beta\partial\gamma:\ .
\ee 
We define the total ghost stress tensor and supercurrent as
\be T^{gh}(z)=T_{bc}(z)+T_{\beta\gamma}(z)
\ee
\be G_{gh}(z)=-(\partial\beta) c(z)-\frac{3}{2}\beta\partial c(z)  -2b\gamma(z).
\ee
The $\beta,\gamma$ system has an NS and an R sector, so that the ghost sector is a direct sum $\CH^{gh}=\CH^{gh}_{\rm NS}\oplus \CH^{gh}_{\rm R}$. Each $\CH^{gh}_{\rm NS}$ and $\CH^{gh}_{\rm R}$  which should be tensored with the corresponding (NS or R) sector of the matter SVA. It is often useful to rewrite the superghosts as
\be \beta=\partial \xi e^{-\phi}\qquad \gamma=\eta e^{\phi}\ ,
\ee
in terms of two fields $\xi$ and $\eta$ of conformal weights $0$ and $1$, respectively, satisfying the same OPEs as the $b$ and $c$ fields, and of a chiral scalar $\phi$.  The latter field always appears either in the form $\partial^n\phi$, $n>0$ or $e^{\frac{m}{2}\phi}$, $m\in \ZZ$. As for the matter theory, the $\CH_{\rm NS+}^{gh},\CH_{\rm NS-}^{gh},\CH_{\rm R+}^{gh},\CH_{\rm R-}^{gh}$ sector are characterized by $m$ being congruent to $0,1,2$ and $3\mod 4$, respectively. In terms of the fields $\phi,\xi,\eta$, the superghost energy momentum tensor $T_{\beta\gamma}$ is
\be T_{\beta\gamma}=-\frac{1}{2}\partial\phi\partial\phi-\partial^2\phi-\eta\partial\xi\ .
\ee
The ghost vertex algebras $\CH^{gh}$ correspond to the ``restricted'' vertex algebra generated by $b,c,\eta,\partial\xi,\phi$, i.e. not including the field $\xi$ without derivatives.

\subsection{BRST cohomology and physical states}

We can now construct, following the standard string theory prescriptions, the chiral analogue of the complete worldsheet SCFT by tensoring the ghost and matter sectors together and implementing a GSO projection. We provide some additional details in Appendix \ref{a:algebradetails}. The resulting GSO projected vertex algebra is
\be \nonumber \CH_{\rm GSO}=(\CH^{m}_{\rm NS+}\otimes \CH^{gh}_{\rm NS+})\oplus (\CH^{m}_{\rm NS-}\otimes \CH^{gh}_{\rm NS-})\oplus (\CH^{m}_{\rm R+}\otimes \CH^{gh}_{\rm R+})\oplus (\CH^{m}_{\rm R-}\otimes \CH^{gh}_{\rm R-}).
\ee

Let us define the BRST charge $Q$, namely the zero mode of the BRST current
\be j_{\rm BRST}(z)=cT(z)+\gamma G(z) +\frac{1}{2}(cT^{gh}(z)+\gamma G^{gh}(z))
\ee
and the ghost and picture numbers $j_0^N$ and $j_0^P$, where
\be j^N=-:\xi\eta:+:bc:,\qquad\qquad j^P=-\partial\phi+:\xi\eta:\ .
\ee  
The $L_0$ eigenvalue, the ghost and picture numbers of the generators of $V_{GSO}$ are as follows
\begin{center}\begin{tabular}{c|ccc}
		& $L_0$ & $j_0^N$ & $j_0^P$ \\ \hline
		$e^{n\phi}$ & $-\frac{1}{2}n(n+2)$ & $0$ & $n$\\ 
		$\partial\xi$ & $1$ & $-1$ & $1$\\
		$\eta$ & $1$ & $1$ & $-1$\\
		$b$ & $2$ & $-1$ & $0$\\
		$c$ & $-1$ & $1$ & $0$
\end{tabular}\end{center}
Notice that $Q$ increases the ghost number but commutes with the picture number. 

One defines the $Q$-cohomology on the space
\be B=\CH_{\rm GSO}\cap \ker b_0\ .
\ee One can split $B$ as
\be B=\oplus_{k \in \Gamma^{1,1}} B(k)=\oplus_{k,h,p,n} B(k)^h_{p,n}
\ee
where $k\equiv(k^0,k^1)\in \Gamma^{1,1}$ is momentum in $1+1$ space-time directions (i.e. the eigenvalue of the zero modes $(\alpha_0^0,\alpha_0^1)$ of $\partial X^\mu$, $h$ is the $L_0$-eigenvalue, and $p$ and $n$ are the $j_0^p$ and $j_0^N$ eigenvalues. Since we will soon pass to the space of physical states, and therefore want to focus on the $L_0=0$ subspace, we also define
\be C(k)_{p,n}=B^0_{p,n}\qquad C(k)=\oplus_{p, n} C(k)_{p,n}\qquad C=\oplus_k C(k)\ .
\ee Then, for each picture number $p$ and momentum $k$, one can define a complex
\be \ldots\stackrel{Q}{\longrightarrow}C(k)_{p,n-1}\stackrel{Q}{\longrightarrow}C(k)_{p,n}\stackrel{Q}{\longrightarrow}C(k)_{p,n+1}\stackrel{Q}{\longrightarrow}\ldots
\ee and the corresponding cohomology groups $H(k)_{p,n}$.

The main results regarding the spaces $H(k)_{p,n}$ follow from \cite{Sch1}, which we summarize here for later use:
\begin{itemize}
	\item There is a non-degenerate bilinear form $(,)_H$ on $H$ induced by an invariant non-degenerate bilinear form $(,)$ on $\CH_{GSO}$ by means of an intermediate form $(,)_C$ on $C$. It is defined by
	\be ([u],[v])_{H}=(u,v)_C=(c_0u,v)\ .
	\ee This form pairs $H(k)_{p,n}$ with $H(-k)_{p',n'}$, where $p+p'=-2$ and $n+n'=2$. (See Propositions 5.2, 5.4, 5.5 of \cite{Sch1}). 
	\item The picture changing operator $X=(Q\xi)_0=\{Q,\xi_0\}$ provides a homomorphism
	\be X:H(k)_{p,n}\to H(k)_{p+1,n}
	\ee which is an isomorphism if $k\neq 0$. The inverse is furnished by (a scalar multiple of) a zero mode of the momentum operator (Proposition 5.8 of \cite{Sch1}). This means that, at least for $k\neq 0$, one can restrict to study the canonical pictures $p=-1$ for the NS sector and $p=-1/2$ for the R sector. The $p=-3/2$ picture is also useful for the R sector, since the bilinear form $(,)_H$ pairs $H(k)_{-1/2,1}$ with  $H(-k)_{-3/2,1}$.

\end{itemize}
Physical states in superstring theory are then given by the BRST cohomology 
\be \bigoplus_{k\in \Gamma^{1,1}} (H_{-1,1}(k)\oplus H_{-1/2,1}(k))\ ,
\ee
with ghost number $1$ and canonical picture numbers $-1$ (for the NS sector) and $-1/2$ (for the Ramond sector). In the GSO projected vertex algebra, these picture numbers correspond to the sectors
\be \CH^{m}_{\rm NS-}\otimes \CH^{gh}_{\rm NS-}
\ee and
\be \CH^{m}_{\rm R-}\otimes \CH^{gh}_{\rm R-}\ .
\ee

\subsection{A Lie superalgebra of physical states}

In this section, we will show, following \cite{Sch1} and \cite{LZ}, that one can define the structure of a Lie superalgebra on the space of physical states in this superstring theory.
The first step is to introduce a bracket $\{,\}:C(k)_{p,n}\times C(k')_{q,m}\to C(k+k')_{p+q,n+m-1}$ (introduced by Lian and Zuckerman \cite{LZ})  given by
	\be \{u,v\}=(-1)^{|u|}(b_{-1}u)_0v\ ,
	\ee where $|u|\equiv n+2p\mod 2\in \ZZ/2\ZZ$ is the parity of $C(k)_{p,n}$. Note that $u$ has $L_0$ eigenvalue $0$ (since it is in $C$), so $b_{-1}u$ is a current. The bracket satisfies the $\ZZ_2$-graded symmetric properties and Jacobi identity
	\be\nonumber \{u,v\}+(-1)^{(|u|+1)(|v|+1)}\{v,u\}=0
	\ee
	\hspace{-.5in}\be\nonumber (-1)^{(|u|+1)(|w|+1)}\{u,\{v,w\}\}+(-1)^{(|v|+1)(|u|+1)}\{v,\{w,u\}\}+(-1)^{(|w|+1)(|v|+1)}\{w,\{u,v\}\}=0
	\ee Furthermore, it is compatible with the actions of the spectral flow and BRST operators, i.e.
	\be X\{u,v\}=\{Xu,v\}=\{u,Xv\}\ 
	\ee and
	\be Q\{u,v\}=\{Qu,v\}=\{u,Qv\}\ 
	\ee
	 (Propositions 5.13, 5.14, 5.15 in \cite{Sch1}). The latter property ensures that $\{,\}$ induces a well-defined bracket (which we denote by the same symbol) $\{,\}:H(k)_{p,n}\times H(k')_{q,m}\to H(k+k')_{p+q,n+m-1}$ on the BRST cohomology classes.

Given these ingredients, one can define a $\Gamma^{1,1}$-graded Lie super-algebra $\tilde\g$, with even part $\tilde\g_0=\oplus_k \g_0(k)$ and odd part $\tilde\g_1=\oplus_k \g_1(k)$, where \be \g_0(k)=H(k)_{-1,1}\qquad\g_1(k)=H(k)_{-1/2,1}\ ,\ee and Lie bracket
	\be [u,v]:=X\{u,v\}\qquad u \in \g_0 \text{ or } v\in \g_0,
	\ee
	\be [u,v]:=\{u,v\}\qquad u, v \in \g_1.
	\ee
	In order to interpret the Lie superalgebra that we obtained as a Borcherds-Kac-Moody algebra, we would like to define an invariant non-degenerate bilinear form on $\tilde \g$. 
	
	For the even part, we can simply take the invariant form on $H$, since it pairs $\g_0(k)=H(k)_{-1,1}$ with $\g_0(-k)=H(-k)_{-1,1}$ non-degenerately. The odd part is more complicated since $(,)_H$ pairs $\g_1(k)=H(k)_{-1/2,1}$ to $H(-k)_{-3/2,1}$. For $k\neq 0$, we can use the isomorphism given by the picture changing operator $X:H(k)_{-3/2,1}\to H(k)_{-1/2,1}$ to identify these two spaces. However, this does not work for $k=0$, since $X$ is not an isomorphism. 
	
	In fact, a direct calculation analogous to the one in Proposition 5.12 of \cite{Sch1} shows that $XH(0)_{-3/2,1}=0$. So, a non-degenerate bilinear form can only be defined on the subspace $\g\subset \tilde \g$ obtained by excluding $\g_1(0)$: the odd, zero-momentum part. 
	It turns out that $\g$ is a genuine subalgebra of $\tilde \g$ and therefore closed under commutation. To see this, note that an element in $\g_1(0)$ can be written as a commutator $[x,y]$ only
	if $x\in \g_0(k)=H_{-1,1}(k)$ and $y\in \g_1(-k)=H_{-1/2,1}(-k)$ (or vice versa). But by definition $[x,y]=X\{x,y\}$ where $\{x,y\}\in H_{-3/2,1}(0)$; since, as already mentioned, $XH_{-3/2,1}(0)=0$, we obtain $[x,y]=0$.
	This implies that no element of $\g_1(0)$ can be obtained as a commutator of two elements in $\tilde\g$, and the subspace $\g$ obtained by excluding $\g_1(0)$ from $\tilde \g$ is indeed a subalgebra of $\tilde \g$.
	
	  We can now define our Borcherds-Kac-Moody algebra $\g$:
	\be \g=\g_0\oplus\g_1\qquad \g_0=\oplus\g_0(k)\qquad \g_1=\oplus_{k\neq 0}\g_1(k)
	\ee with a bilinear form $(,)_\g$ given by
	\be (u,v)_\g=(u,v)_H\qquad \text{for }u,v\in \g_0\ ,
	\ee
	\be (u,v)_\g=-(X^{-1}(u),v)\qquad \text{for }u,v\in \g_1\ ,
	\ee and $(u,v)_\g=0$ in the other cases. 
	
	The algebra $\g$ is, by construction, a Lie superalgebra with a supersymmetric, invariant, non-degenerate bilinear form (see appendix \ref{a:BKM} for more details). In fact, using the characterization given in Theorem 2.5.4 of \cite{Ray}, it can be easily shown that $\g$ is actually a Borcherds-Kac-Moody (BKM) superalgebra as claimed. The definition and the main properties of BKM superalgebras, as well as the proof that $\g$ satisfies the characterization of \cite{Ray}, are given in Appendix \ref{a:BKM}.

\subsection{Root multiplicities}\label{sec:mult}

In this section, we calculate the dimensions of the spaces $\g_0(k)=H(k)_{-1,1}$ and $\g_1(k)=H(k)_{-1/2,1}$ and use this result to compute the denominator of the algebra $\g$. For $k^2=0$, we can compute the dimensions of these spaces directly by finding a basis of $Q$-closed states and modding out by the $Q$-exact states. The procedure is very similar to \cite{Sch1} and some details are reported in Appendix \ref{a:zeromomentum}. We just report the final results for for picture number $-1/2$ (i.e. the odd roots):
\be\label{rootdim}
\begin{matrix}
	& \dim H(k)_{-1/2,1} \\
	k^0=-k^1\neq 0	& 0 \\
	k^0=k^1	& 24 
\end{matrix}
\ee
and for picture number $-1$ (i.e. even roots)
\be
\begin{matrix}
	& \dim H(k)_{-1,1} \\
	k^0=\pm k^1,\ k\neq 0	& 0 \\
	k=0 & 2
\end{matrix}
\ee

 The $24$ generators $\CQ_i$, $i=1,\ldots,24$, of $H(0)_{-1/2,1}$ represent $24$ spacetime supercharges (cf. \cite{PPV, PPV2} and especially \cite{Distleretal}), while the two generators $P^\mu$, $\mu=0,1$, of $H(0)_{-1,1}$ represent the spacetime momentum in $1+1$ dimensions. The Lie superalgebra $\g_0(0)\oplus\g_1(0)\subset \tilde \g$ generated by these $k=0$ components is an $\CN=(24,0)$ supersymmetry algebra in $1+1$ dimensions, with (anti-)commutation relations
\be \{\CQ_i,\CQ_j\}=2\delta_{ij}(P^0+P^1)\ ,\qquad [P^\mu,P^\nu]=0=[P^\mu,\CQ_i]\ .
\ee As described above, the states in $H(0)_{-1/2,1}$ are included in the Lie algebra $\tilde \g$, but not in the BKM algebra $\g$, where the non-degenerate bilinear form is defined. 

In spite of that, note that the graded components $\g(k)\subset \g$ for $k\neq 0$ are still \textit{modules} for the $\CN=(24,0)$ supersymmetry algebra above.  In particular, the component $\g(k)$ is the eigenspace for $P^0,P^1$ with eigenvalues $k^0,k^1$. For $k^2\neq 0$, one has $k^0+k^1\neq 0$, so that the adjoint action ${\rm ad}_{\CQ_i}$ of each supercharge $\CQ^i\in \g_1(0)$, $i=1,\ldots,24$, on $\g(k)$ squares to $({\rm ad}_{\CQ_i})^2=(k^0+k^1){\rm id}$. In particular, it defines an invertible map from the odd component $\g_1(k)$ to the even component $\g_0(k)$. We conclude that
\be \dim\g_0(k)=\dim\g_1(k)\qquad \text{for all }k,\ k^2\neq 0\ .
\ee
Therefore, for $k^2\neq 0$ it is sufficient to compute the dimension of the even part $\g_0(k)$.

The simplest way to do this is to use the fact, proved in \cite{LZ} (see proposition 5.6 in \cite{Sch1}), that for $k\neq 0$ the cohomology space $H(k)_{-1,n}=0$ for $n\neq 1$. The Euler-Poincar\'e principle gives
\be -\dim H(k)_{-1,1}=\sum_n (-1)^n\dim C(k)_{-1,n}\ .
\ee In order to compute the right-hand side of this equation one can take the $q^0$ coefficient in the partition function
\be Z_k(\tau)=\sum_{h,n} q^h (-1)^n\dim B^h_{-1,n}(k)=\Tr_{B_{-1}(k)}((-1)^{j_0^N}q^{L_0})\ .
\ee The latter trace is the product of the contributions coming from the internal CFT $V^{f\natural}$ (with negative fermion number, due to GSO projection), the spacetime bosons and fermions, the ghosts, and the superghosts. Notice that for $k\neq 0$, the computation can be performed in any picture number, since the spaces $H(k)_{p,1}$ for any $p\in \ZZ$ are isomorphic via the picture changing operator. It is a standard result in superstring theory that the contributions from the ghosts\footnote{Note that the mode $c_0$ does not contribute, because we are considering only states in $\ker b_0$.} $b,c$
\be (1-q^{-1})\prod_{n=1}^\infty (1-q^n)\prod_{n=2}^\infty(1-q^n)=-q^{-1}\prod_{n=1}^\infty (1-q^n)^2
\ee and from the superghosts $\beta,\gamma$ 
\be \frac{1}{\prod_{n=1}^\infty (1+q^{n+\frac{1}{2}})(1+q^{-1/2})\prod_{n=1}^\infty (1+q^{n-\frac{1}{2}})}=\frac{1}{q^{-1/2}\prod_{n=1}^\infty (1+q^{n-\frac{1}{2}})^2}\ ,
\ee
cancel almost exactly (up to a factor $-q^{-1/2}$) against the bosonic and fermionic oscillators $\psi^\mu_r$, $\alpha^\mu_n$
\be \frac{\prod_{n=1}^\infty (1+q^{n-\frac{1}{2}})^2}{\prod_{n=1}^\infty (1-q^n)^2}
\ee
in the light-cone directions $\mu=0,1$ --- in our case, these are all of the space-time directions, since we only have $1+1$ dimensions. What remains is the contribution $q^{1/2}F_{\rm NS-}(\tau)$ from the internal CFT and $q^{k^2/2}$ from space-time momentum, so that
\be \Tr_{B_{-1}(k)}((-1)^{j_0^N}q^{L_0})=-q^{k^2/2}F_{\rm NS-}(\tau)\ ,
\ee and the $q^0$ term is just the Fourier coefficient $-C_{\rm NS-}(-k^2/2)$ of $-F_{\rm NS-}$. Taking the signs into account, we get
\be \dim \g_0(k)=C_{\rm NS-}(-\frac{k^2}{2})\ .
\ee

For $k\neq 0$, the same result can be obtained by invoking the no-ghost theorem \cite{Goddard:1972iy,Polchinski:1998rr}, which guarantees that the BRST quantization is equivalent to the light-cone quantization. In particular, the BRST cohomology space $H(k)_{-1,1}$ is isomorphic to the component of momentum $k$ in the `transverse space' of the matter CFT, i.e. the subspace of the matter CFT with no bosonic or fermionic oscillators in the $1+1$-dimensional space-time and satisfying the ``mass-shell'' condition $L_0-1/2=0$. This subspace is spanned by states of the form $|v\rangle\otimes |k\rangle\in V_{\rm NS-}\otimes V_{1,1}$, where $|k\rangle\in V_{1,1}$ is the state corresponding to the field $e^{ikX}$, $k\in \Gamma^{1,1}$, and  $|v\rangle \in V_{\rm NS-}$ is a state of negative fermion number in $V_{\rm NS}\cong V^{f\natural}$ with $L_0$ eigenvalue $h\in 1/2+\ZZ$, satisfying the mass shell condition $h=\frac{k^2}{2}+\frac{1}{2}$. This gives an isomorphism
\be \g_0(k)\cong V_{\rm NS-}(-\frac{k^2}{2})\ ,\qquad k\neq 0\ ,
\ee where $V_{\rm NS\pm}(n)$ and $V_{\rm R\pm }(n)$ denote the subspaces of $V_{NS\pm}$ and $V_{R\pm}$ with $L_0$ eigenvalue $n+\frac{1}{2}$.
As mentioned above, for $k$ of non-zero norm ($k^2\neq 0$), one has $\dim \g_0(k)=\dim \g_1(k)$, while for $k\neq 0$ of norm zero, one has $\dim\g_1(k)=24$ if $k^0=k^1$ and $\dim\g_1(k)=0$ if $k^0=-k^1\neq 0$. In fact, for $k^0=k^1$, $\g_1(k)$ is naturally isomorphic to the $24$-dimensional ground state space $V_{\rm R-}(0)$ of $V_{\rm R-}$, which in turn is the subspace of $V_{\rm R}\cong V^{f\natural}_{tw}$ with negative fermion number. On the other hand, $\g_1(k)$ for $k^2\neq 0$ can be obtained by commutation of $\g_1(0)$ with $\g_0(k)$ and is therefore isomorphic to the subspace of $V_{\rm R-}(-k^2/2)$ with negative fermion number and $L_0$ eigenvalue $-k^2/2+1/2$. To summarize, we have
\be \g_1(k)\cong V_{\rm R-}(-\frac{k^2}{2})\ ,\qquad k^0\neq -k^1
\ee so that
\be \dim \g_1(k)=C_{\rm R-}(-\frac{k^2}{2})\ ,\qquad k^0\neq -k^1
\ee and $\dim \g_1(k)=0$ for $k^0=-k^1$.

Since 
\be \g(k) = \g_0(k)\oplus\g_1(k) = V_{\rm NS-}(-\frac{k^2}{2})\oplus V_{\rm R-}(-\frac{k^2}{2}) = V_{tw}^{s\natural}(-\frac{k^2}{2})\ee we immediately see that our Lie algebra carries a faithful action of $Co_0$: the action inherited from the canonically twisted Conway module $V_{tw}^{s\natural}$. In the next subsection we will describe how this leads immediately to a family of twisted denominator identities for any conjugacy class $[g] \in Co_0$.
\subsection{Denominator}\label{sec:denom}
Given our explicit construction of the BKM algebra and our knowledge of the dimensions of the graded subspaces we can read off the product side of the famous Weyl denominator formula associated to the BKM. We will rewrite the denominators in terms of equivariant Hecke operators in \S \ref{sec:proof} and subsequently prove the additive side of the Weyl denominator formula. In \S \ref{sec:simple} we determine the simple roots of the BKM.  
 
The algebra $\g$ has a Cartan subalgebra $\mathfrak{h}=\g_0(0)$ of rank $2$, generated by $P^0,P^1$. The component $\g(k)$ is the simultaneous eigenspace for $P^0,P^1$ with eigenvalues (root) $k=(k^0,k^1)$.  An element $h\in \mathfrak{h}$ is called regular if $h_\mu k^\mu\neq 0$ for all non-zero roots $k\neq 0$ with $\g(k)\neq 0$. Given a regular element $h$, the positive (respectively, negative) roots are the roots $k$ for which $h_\mu k^\mu>0$ ($h_\mu k^\mu<0$).  Since there are no non-zero roots with $k^0=0$, and we can define the set of positive roots of $\g$ as the ones satisfying $k^0>0$. This yields a triangular decomposition of the algebra
\be \g=\g_-\oplus \mathfrak{h}\oplus \g_+\ ,
\ee where $\g_{\pm}=\oplus_{\pm k^0>0}\g(k)$ and $\mathfrak{h}=\g(0)$.
The algebra denominator is given by \cite{Ray}
\be \prod_{\substack{k\in \Gamma^{1,1}\\k^0>0}}(1-e^{k\cdot  \zeta})^{\dim \g_0(k)}(1+e^{k \cdot \zeta})^{-\dim \g_1(k)}
\ee and the superdenominator by
\be \prod_{\substack{k\in \Gamma^{1,1}\\k^0>0}}(1-e^{k\cdot \zeta})^{\dim \g_0(k)}(1-e^{k\cdot  \zeta})^{-\dim \g_1(k)}\ .
\ee
 We can take a basis of the lattice $\Gamma^{1,1}$ so that $k=(m,n)\in \ZZ\oplus\ZZ$ and $\zeta=(2\pi i\tau,2\pi i\sigma)\in \Gamma^{1,1}\otimes\CC\cong \CC\oplus\CC$, with $k\cdot \zeta=2\pi i(m\sigma+n\tau)$. In this basis, the $k^0$ and $k^1$ components of $k$ are given by
 \be\label{windmom} k^0=-k_0=\frac{1}{\sqrt{2}}\left(\frac{m}{R}+nR\right),\qquad k^1=k_1=\frac{1}{\sqrt{2}}\left(\frac{m}{R}-nR\right)\ ,
 \ee for some positive real number $R$, and the norm is
 \be k^2=-(k^0)^2+(k^1)^2=-2mn\ .
 \ee 
In the string theory picture we will describe in the next subsection, $R$ will be the radius of the spatial circle in the 1+1 dimensional spacetime.

As discussed above, the even and odd roots with momentum $k=(m,n)$ have multiplicities $C_{\rm NS-}(mn)$ and $C_{\rm R-}(mn)$, respectively. Since $F_{\rm NS-}$ and $F_{\rm R-}$ have no polar terms, there are no roots with $mn<0$ (real roots). The positive roots must satisfy  the condition
 \be k^0=\frac{m}{R}+nR>0\ .
 \ee Positive roots of negative norm ($mn>0$) are given by $m>0$ and $n>0$. By \eqref{rootdim}, roots of zero norm ($mn=0$) only exist for $k\neq 0$ and $k^0=k^1$, which means $n=0$. Therefore, positive roots of zero norm must have $m>0$. We conclude that a root $(m,n)$ is positive if and only if  $m>0$. 
Using these results, we find that the denominator of the BKM superalgebra is
\be\label{denom} \prod_{\substack{m,n\in \ZZ\\m>0}}(1-p^mq^n)^{C_{\rm NS-}(mn)}(1+p^mq^n)^{-C_{\rm R-}(mn)}
\ee
and the superdenominator is
\be\label{superdenom} \prod_{\substack{m,n\in \ZZ\\m>0}}(1-p^mq^n)^{C_{\rm NS-}(mn)}(1-p^mq^n)^{-C_{\rm R-}(mn)}
\ee where we let $p=e^{2\pi i\sigma}$.

The Weyl denominator can be interpreted as a character for the virtual $\Gamma^{1,1}$-graded and $\ZZ_2$-graded space $\bigwedge(\g_+)$, which is the alternating sum
\be \bigwedge(\g_+)=\bigwedge\nolimits^0\g_+ \ominus \bigwedge\nolimits^1\g_+\oplus \bigwedge\nolimits^2\g_+\ominus \bigwedge\nolimits^3\g_+\oplus\ldots \ee 
where $\bigwedge^i \g$ is the exterior power of $\g_+$, i.e. the space spanned by elements of the form $u_1\wedge \ldots\wedge u_i$. We emphasize that, $\g_+$ being a superspace, the wedge product $u\wedge v$ of two elements $u,v\in\g+$ is $\ZZ_2$-graded skew-symmetric, i.e. satisfies $u\wedge v=-(-1)^{|u||v|}v\wedge u$, where $|u|,|v|\in \ZZ_2$ are the parity of $u$ and $v$. 
The Borcherds-Kac-Weyl denominator formula, which will  discuss in \S \ref{sec:simple},  admits a homological interpretation $\bigwedge(\g_+) = H(\g_+)$, where $H(\g_+)$ is the alternating sum of Lie algebra homology groups of $\g_+$. The $H(\g_+)$ side coincides with the sum over Weyl group elements on the additive side of the Weyl denominator formula, equation \ref{eq:denom} (see \cite{GarlandLepowsky} for the Kac-Moody case and \cite{BorcherdsMM,jurisich1998generalized} for the BKM case). The denominator formula then arises from identifying the graded characters of both sides. Because of the identification $\g\cong V^{s\natural}_{tw}$, both $\bigwedge(\g_+)$ and $H(\g_+)$ carry a natural action of the Conway group $Co_0$. Following  \cite{BorcherdsMM}, one can therefore obtain a twisted denominator formula for each element $g \in Co_0$ (or rather conjugacy class in $Co_0$)  by equating the ``$g$-twisted characters'' of $\bigwedge(\g_+)$ and $H(\g_+)$, i.e.  replacing the virtual graded dimensions in the characters with graded traces of $g$. 
We will write down the twisted denominator formulae explicitly in \S \ref{sec:twisted}. 

\subsection{The BKM superalgebra in type II string theory}\label{sec:strings}

In the previous subsections, we constructed a BKM superalgebra in a chiral version of superstring theory, where only holomorphic fields where present. This is consistent only upon compactifying all space-time directions -- including time -- in a suitable holomorphic superconformal field theory. This setup  is very similar to Scheithauer's construction of a superalgebra in \cite{Sch1}, and is the closest analogue to Borcherds' original  construction of the Monstrous Lie algebra \cite{BorcherdsMM}, where a chiral version of bosonic string theory was used, where all space-time directions were compactified in a bosonic conformal field theory.

While this construction is formally consistent, it would be desirable to find an interpretation of this algebra in a more standard non-chiral string theory describing a physically meaningful space-time where at least the time direction is uncompactified.\footnote{We note here that with a periodic time-like direction, which is required in order to get a chiral superstring theory that is  formally consistent from the world-sheet perspective, some of the basic requirements of a physical theory, such as causality, would not make sense.} In \cite{PPV} and \cite{PPV2} it was noticed that Borcherds' Monstrous Lie algebra can be also obtained as some kind of algebra of BPS states (in a suitable sense) in a particular low dimensional compactification of a heterotic string theory. This suggests that the BKM superalgebra we just considered might admit a similar interpretation as a BPS algebra in some fully fledged non-chiral superstring  compactification. In this section we will sketch a possible route toward the definition of this BPS algebra; we leave the details to future work.

Let us focus, for definiteness, on the type IIA superstring; a similar construction should apply in type IIB. As a first step, we compactify the ten dimensional type IIA superstring on the superconformal field theory $V^{f\natural}\otimes \overline{V^{f\natural}}$ (which we will call the internal CFT), where $\overline{V^{f\natural}}$ is an antiholomorphic (right-moving) version of the SVOA $V^{f\natural}$. More precisely, $V^{f\natural}\otimes \overline{V^{f\natural}}$ represents the NS-NS sector of the theory, while the R-NS, NS-R and R-R sectors are obtained by a suitable replacement of $V^{f\natural}$ or $\overline{V^{f\natural}}$ by the corresponding twisted module. We obtain a model  with $1+1$-dimensions in space-time and $\CN=(24,24)$ supersymmetry. Notice that the GSO projection relates the fermion number in the internal CFT with chirality in space-time. Next, we further compactify the spatial direction on a circle of radius $R$. We indicate with $k^0_L,k^0_R$ and $k^1_L,k^1_R$ the left- and right-moving momenta in the time and space directions, respectively. Since the time direction is uncompactified, we get
\be\label{matchmom} k^0_L=k^0_R\ ,
\ee while the possible momenta $(k^1_L,k^1_R)$ in the space direction form an even unimodular (Narain) lattice $\Gamma^{1,1}$ with quadratic form $(k^1_L)^2-(k^1_R)^2$. $(k^1_L,k^1_R)$ are related to momentum $m$ and winding number $n$ along $S^1$ as
\be\label{momlattice} k^1_L=\frac{1}{\sqrt{2}}(\frac{m}{R}-nR),\qquad k^1_R=\frac{1}{\sqrt{2}}(\frac{m}{R}+nR)\ .
\ee

The physical states are given by the cohomology with respect to a nilpotent BRST charge $Q=Q_L\oplus Q_R$, where $Q_L$ and $Q_R$ are, respectively, the BRST charge in the chiral compactification described in the previous subsections, and ts anti-holomorphic analogue. The physical states are then obtained by tensoring in a suitable way some holomorphic  $Q_L$-cohomology spaces $H_L(k_L)_{-1,1}$ and $H_L(k_L)_{-1/2,1}$,  with suitable right-moving $Q_R$-cohomology spaces  $H_R(k_R)_{-1,1}$ and $H_R(k_R)_{-1/2,1}$. To avoid confusion, we will refer to the BRST space obtained in the previous subsections as the `chiral' BRST spaces.  We want to focus on the space of $1/2$-BPS physical states (i.e. BRST classes) that are invariant under the $24$ space-time supercharges $\CQ^i_R$ coming from the NS-R sector. These $24$ supercharges generate a two-dimensional supersymmetry algebra 
\be \{\CQ^i_R,\CQ^j_R\}=2(P^0_R-P^1_R)\delta^{ij}\ ,
\ee
where $P^0_R$ and $P^1_R$ are the right-moving momenta with eigenvalues $k^0_R, k^1_R$. As a consequence, a physical state is annihilated by all $\CQ^i_R$ if and only if its right-moving momenta  satisfy
\be\label{BPS} k^1_R=k^0_R\ .
\ee This BPS condition, together with \eqref{matchmom} and \eqref{momlattice}, implies that the left-moving momenta $k_L\equiv (k^0_L,k^0_R)$ take values in the $\Gamma^{1,1}$ lattice and are related to the winding $n$ and momentum $m$ as in \eqref{windmom}. Thus, the space of BPS physical states for $k_R\neq 0$  is obtained by tensoring the left-moving spaces space $H_L(k_L)_{-1,1}$ or  $H_L(k_L)_{-1/2,1}$, which are isomorphic to the chiral BRST cohomology spaces $H(k_L)_{-1,1}$ and   $H(k_L)_{-1/2,1}$, with the space $H_R(k_R)_{-1/2,1}$, which is isomorphic to the $24$-dimensional ``chiral'' space $H(k_R)_{-1/2,1}$ for $k^1_R=k^0_R\neq 0$ (recall that $H(k_R)_{-1,1}=0$ for $k^0_R=k^1_R\neq 0$).  Thus, there are $24$ BPS classes for each element in the algebra $\g(k)$, $k\neq 0$, and we denote them schematically as $v\otimes \chi^i$, where $v\in \g(k_L)$ and $\chi^i\in H(k_R)_{-1/2,1}$, with $k_L$ and $k_R$ related as above. The positive roots then correspond to BPS physical states of positive energy $k^0_L=k^0_R>0$. 

Let us now choose a representative (that we still denote by $v\otimes\chi^i$, with some abuse of notation) for each of these BPS classes. Note that the right-moving part $\chi^i$ of this representative is uniquely defined, since there are no BRST exact states in $C(k_R)_{-1/2,1}$ for $(k_R)^2=0$. By acting on this state with one of the space-time supercharges $\CQ^i_R$, one obtains either $0$ or a BRST exact state.   In fact, for each such class there is exactly one supercharge $\CQ^i_R$ which yields a non-zero BRST exact state, and the right-moving part of this state is of the form $k_{\mu,R} \tilde\psi^\mu_{-1/2}e^{-\tilde\phi}_{-1/2}|k_R\rangle$. In particular, this state depends only on $k_R$, and not on which of the $24$ states $\chi^i$ in $H(k_R)_{-1/2,1}$ we are considering.\footnote{With some abuse of notation, we are not distinguishing between classes in  $\tilde{H}(k_R)_{-1/2,1}$ and their representatives; this is justified because there are no BRST exact states in $\tilde{C}(k_R)_{-1/2,1}$.} Therefore, one can associate with each element in the algebra $\g$ a BRST exact ``BPS state'' in the type II superstring algebra.\footnote{More precisely, there is a BRST exact state for each choice of the representative of a BRST class satisfying the BPS condition. This exact state is the $Q_R$ image of a $Q_L$ closed state. The one-to-one correspondence exists between the algebra $\g$ and the quotient of the space of these $Q_R$-exact, $Q_L$-closed states modulo its subspace of $Q_L$-exact states.} We conjecture that, in analogy with \cite{PPV} and \cite{PPV2}, natural (anti-)commutation relations can be defined among these BRST exact states so that they form a superalgebra  isomorphic to $\g$. Note that the right-moving BRST exact state belongs to the right-moving NS sector (since it comes from acting by a supercharge on a Ramond state), so that the elements in $\g_0$ are associated with space-time bosons in the  NS-NS sector and the elements in $\g_1$ are associated with space-time fermions in the  R-NS sector.

In this sense, the superalgebra $\g$ can be interpreted as an algebra of BPS states in the type II superstring model.

\section{Hecke operators and the denominator identity}\label{sec:proof}
In this section, we will show that the infinite products defining the denominator and superdenominator of the BKM superalgebra $\g$ are (up to a simple factor) modular forms in $\sigma$ and $\tau$. We will use this fact to obtain new expressions for these denominators, which will be interpreted in \S \ref{sec:simple} as the additive side of the Borcherds-Kac-Weyl denominator formula. 

We would like to rewrite the product side of the denominator formula using the machinery of (equivariant) Hecke operators. 
We will build up to the definition of the equivariant Hecke operators acting on a vector-valued modular function $f$:

\be f(x,y;\tau)=\sum_{n\in \ZZ} c_{x,y}(\frac{n}{N})q^{\frac{n}{N}}\ ,\ee with $x,y\in \ZZ/N\ZZ$, satisfying
\be\label{ftransf} f\left(x,y;\frac{a\tau+b}{c\tau+d}\right)=f(ax+cy,bx+dy;\tau)\qquad \left(\begin{smallmatrix}
	a & b\\ c & d
\end{smallmatrix}\right)\in SL(2,\ZZ)\ .
\ee 
One can also use a discrete Fourier transform to define the functions
$F_{r,s}$, $r,s\in \ZZ/N\ZZ$ by
\be\label{Fdef} F_{d,r}(\tau)=\sum_{n}C_{d,r}(n)q^n=\frac{1}{N}\sum_{k\in \ZZ/N\ZZ} e^{-2\pi i\frac{kr}{N}} f(d,k;\tau)\ .
\ee
In our setup, the functions of interest are $f(x,y;\tau)$ with $x,y\in \ZZ/2\ZZ$ given by equations (\ref{eq:Vsnat1})-(\ref{eq:Vsnat4}).
Via \eqref{Fdef}, one can then define the four functions
\begin{align} &F_{00}(\tau)=F_{\rm R+}=F_{\rm NS-}=2048q+\ldots
\\& F_{01}(\tau)=-F_{\rm R-}=-24-2048q-\ldots
\\& F_{10}(\tau)=-F_{\rm NS-}=-F_{\rm R+}=-2048q-\ldots
\\&F_{11}(\tau)=F_{\rm NS+}=q^{-1/2}+276 q^{1/2}+\ldots
\end{align}
where we set
\be \label{eq:bigC} F_{d,r}=\sum_n C_{d,r}(n) q^n\ .
\ee
While these are the functions we are ultimately interested in, it is useful to discuss the equivariant Hecke operators for more general $f$.

Let $M$ be the monoid $M:=\oplus_{m=1}^\infty M(m)$ of integral $2\times 2$ matrices of positive determinant, where $M(m)$ is the component with determinant $m$
\be M(m):=\{\left(\begin{smallmatrix}
	a & b\\ c & d
\end{smallmatrix}\right)\mid a,b,c,d\in \ZZ,\ ad-bc=m\}\ .
\ee We can define an action of $M$ on $f$ by
\be (f_{\big \rvert \left(\begin{smallmatrix}
		a & b\\ c & d
	\end{smallmatrix}\right)})(x,y;\tau)=f(dx-cy,ay-bx;\frac{a\tau+b}{c\tau+d})\qquad \left(\begin{smallmatrix}
a & b\\ c & d
\end{smallmatrix}\right)\in M\ .
\ee where $x,y\in \ZZ/N\ZZ$. It is easy to check that
\be f_{ \rvert \gamma_1\gamma_2}=(f_{\rvert \gamma_1})_{\rvert\gamma_2}\qquad \gamma_1,\gamma_2\in M\ .
\ee Now, the transformation properties \eqref{ftransf} imply $f$ is invariant under the action of $SL(2,\ZZ)\equiv M(1)\subset M$
\be\label{sl2zact} (f_{\big \rvert \left(\begin{smallmatrix}
		a & b\\ c & d
	\end{smallmatrix}\right)})(x,y;\tau)=f(x,y;\tau)\qquad \left(\begin{smallmatrix}
a & b\\ c & d
\end{smallmatrix}\right)\in SL(2,\ZZ)\ .
\ee
Under this assumption on $f$, one can define the equivariant Hecke operators
\begin{align*} (T_mf)(x,y;\tau)&=\frac{1}{m}\sum_{[u]\in SL(2,\ZZ)\backslash M(m)} (f_{\rvert u})(x,y;\tau)\\ &=\frac{1}{m}\sum_{\substack{a,d\in \ZZ_{>0}\\ ad=m}}\sum_{ b\mod d} f(dx,ay-bx;\frac{a\tau+b}{d})\ ,
\end{align*}where $u$ is a representative in the coset $[u]\in SL(2,\ZZ)\backslash M(m)$; invariance of $f$ under $SL(2,\ZZ)$ implies that the Hecke operator does not depend on the choice of representative. In the explicit formula, we used the fact that every representative of $SL(2,\ZZ)\backslash M(m)$ can be chosen of the form $\left(\begin{smallmatrix}
a & b\\ 0 & d
\end{smallmatrix}\right)$ where $ad=m$ and $0\le b<d$.
It is easy to see that $T_mf$ is still invariant under the action of $SL(2,\ZZ)$
\be (T_mf)_{\rvert\gamma}=\frac{1}{m}\sum_{[u]\in SL(2,\ZZ)\backslash M(m)} f_{\rvert u\gamma}=\frac{1}{m}\sum_{[u\gamma]\in SL(2,\ZZ)\backslash M(m)} f_{\rvert u\gamma}=T_mf
\ee for all $\gamma\in SL(2,\ZZ)$, where $[u\gamma]\in SL(2,\ZZ)\backslash M(m)$ is the coset corresponding to $u\gamma$.

Notice also that if two functions $f,g:(\ZZ/N\ZZ)\times (\ZZ/N\ZZ)\times \CH\to \CC$ are invariant under the $SL(2,\ZZ)$ action \eqref{sl2zact}, then not only linear combinations
$ af+bg$, $a,b\in \CC$, but even the product $fg$ and powers $f^n$, $n\in \NN$ are invariant under the same $SL(2,\ZZ)$ action. As a consequence, we obtain that the function\footnote{Here, in principle, we consider $\Psi_f$ as a formal power series in $p=e^{2\pi i\sigma}$ with coefficients in functions $(\ZZ/N\ZZ)\times (\ZZ/N\ZZ)\times \CH\to \CC$. Invariance under $SL(2,\ZZ)$ is really invariance of each coefficient. The work by Borcherds \cite{Borcherds:1996uda} shows that this formal power series actually converges absolutely uniformly for $(\sigma,\tau)$ in a suitable domain of $\CH\times\CH$.}
\begin{align} \Psi_f(x,y;\tau):=&(\exp(\sum_{m=1}^{\infty}p^mT_mf ))(x,y;\tau)=\sum_{k=0}^{\infty} \frac{1}{k!}(\sum_{m=1}^\infty p^mT_mf)^k(x,y;\tau)\end{align} is also $SL(2,\ZZ)$ invariant, which implies
\be\label{Psitransf} \Psi_f\left (x,y;\frac{a\tau+b}{c\tau+d}\right)=\Psi_f(ax+cy,bx+dy;\tau)\qquad \left(\begin{smallmatrix}
	a & b\\ c & d
\end{smallmatrix}\right)\in SL(2,\ZZ)\ .
\ee 
The function $\Psi_f$ admits a representation as an infinite product (see appendix \ref{a:prfusefulid} for a proof)
\be\label{infprod} \Psi_f(x,y;\tau)=\prod_{d=1}^\infty \prod_{\substack{t,k\in \ZZ/N\ZZ\\t\equiv ky\pmod N}}\prod_{\substack{r\in\ZZ\\r\equiv kx\pmod N}} (1-e^{2\pi i\frac{t}{N}}p^{d}q^{\frac{r}{N}})^{C_{dx,k}(\frac{rd}{N})}
\ee

Let us now apply this formula to the functions $f(x,y;\tau)$, $x,y\in \ZZ/2\ZZ$, defined in (\ref{eq:Vsnat1})--(\ref{eq:Vsnat4}). In this case (\ref{infprod}) yields the following four infinite product formulas:
\bea \Psi_f(0,0;\tau)&=&\prod_{d=1}^\infty \prod_{k=0}^1\prod_{r\in\ZZ} (1-p^{d}q^{r})^{C_{0,k}({rd})}=\prod_{d=1}^\infty \prod_{r\in\ZZ} (1-p^{d}q^{r})^{C_{0,0}({rd})+C_{0,1}({rd})}
\\\nonumber
&=&\prod_{d=1}^\infty {1\over (1-p^d)^{24}}\\\nn
 \Psi_f(0,1;\tau)&=&\prod_{d=1}^\infty \prod_{k=0}^1\prod_{r\in \ZZ} (1-(-1)^kp^{d}q^{r})^{C_{0,k}(rd)} \\
& =&\prod_{d=1}^\infty \prod_{r\in \ZZ} (1-p^{d}q^{r})^{C_{0,0}(rd)}(1+p^{d}q^{r})^{C_{0,1}(rd)}
\\ \nn
&=&\prod_{d=1}^\infty {1\over (1+p^d)^{24}} {(1-pq)^{2048}\over (1+pq)^{2048}}{(1-pq^2)^{49152}\over (1+pq^2)^{49152}}{(1-p^2q)^{49152}\over (1+p^2q)^{49152}}\cdots\\
\Psi_f(1,0;\tau)&=&\prod_{d=1}^\infty \prod_{r\in\ZZ} (1-p^{d}q^{\frac{r}{2}})^{C_{d,r}(\frac{rd}{2})}\\\nn
&=&{(1-pq^{-{1\over 2}}) (1-pq^{1\over2})^{276}(1-p^3q^{1\over 2})^{11202} (1-pq^{3\over2})^{11202}(1-p^2q)^{49152}\cdots \over (1-p^2q^{1\over 2})^{2048}(1-pq)^{2048} (1-p^4q^{1\over 2})^{49152}\cdots}
\\
 \Psi_f(1,1;\tau)&=&\prod_{d=1}^\infty \prod_{r\in\ZZ} (1-(-1)^rp^{d}q^{\frac{r}{2}})^{C_{d,r}(\frac{rd}{2})}\\\nn
 &=&{(1+pq^{-{1\over 2}}) (1+pq^{1\over2})^{276}(1+p^3q^{1\over 2})^{11202} (1+pq^{3\over2})^{11202}(1-p^2q)^{49152}\cdots \over (1+p^2q^{1\over 2})^{2048}(1-pq)^{2048} (1+p^4q^{1\over 2})^{49152}\cdots}.
\eea
The infinite products $\Psi_f(0,0;\tau)$ and $\Psi_f(0,1;\tau)$ can be recognized as the superdenominator and denominator, respectively, of the BKM superalgebra $\g$ described in \S \ref{sec:denom}. 

On the other hand, we expect $\Psi_f(1,0;\tau)$ and $\Psi_f(1,1;\tau)$ to be the superdenominator and denominator of a distinct BKM algebra which we do not construct here. We will comment more on this point in the coming sections.
\subsection{Proof of the denominator identities}

Using the formalism described in the previous subsection, we will now prove the following set of identities
\bea\label{eq:main0}
p^{-1}\Psi_f(0,0;\t) &=&{1\over \eta^{24}(\sigma)}-24-f(0,0;\t)\\\label{eq:main2}
p^{-1}\Psi_f(0,1;\t) &=&  {\eta^{24}(\sigma)\over \eta^{24}(2\sigma)}+24-f(0,1;\t)\\\label{eq:main1}
p^{-1}\Psi_f(1,0;\t) &=&  f(1,0; 2\sigma)-f(1,0;\t)\\\label{eq:main3}
p^{-1}\Psi_f(1,1;\t) &=&  f(1,0; 2\sigma)-f(1,1;\t).
\eea

 The first two identities can be interpreted as the Borcherds-Weyl-Kac denominator and superdenominator formulae of our BKM, respectively (see \S \ref{sec:simple}). The second two identities should similarly correspond to the denominator and superdenominator formulae of a distinct BKM which we have not constructed. However, we will see once one proves one of (\ref{eq:main2})--(\ref{eq:main3}), the other two immediately follow. 
 
 We begin by considering the identity in (\ref{eq:main0}). Because of the identity
 \be C_{0,0}(n)+C_{0,1}(n)=0 \qquad\qquad \text{for all }n\neq 0\ ,
 \ee the infinite product $p^{-1}\Psi_f(0,0;\t)$ reduces to
 \be p^{-1}\prod_{d=1}^\infty  (1-p^{d})^{C_{0,0}(0)+C_{0,1}(0)}=p^{-1}\prod_{d=1}^\infty  (1-p^{d})^{-24}=\frac{1}{\eta^{24}(\sigma)}\ ,
 \ee and since $f(0,0;\t)=-24$, eq.\eqref{eq:main0} follows immediately.
 
Let us now prove the identities (\ref{eq:main2})--(\ref{eq:main3}); we will use a generalization of Borcherds' proof of Koike-Norton-Zagier identity for the $J$-function \cite{BorcherdsMM}. Using the fact that $f(1,0;\tau)=f(0,1;-1/\tau)= f(1,1;\tau -1)$ and that, by \eqref{Psitransf},
\be \Psi_f(1,0;\tau)=\Psi_f(0,1;-1/\tau)=\Psi_f(1,1;\tau-1)\ ,
\ee it is sufficient to prove one of these identities and the other two follow immediately.

We choose to focus on the identity \eqref{eq:main2}.  By \eqref{Psitransf}, we have
\be \Psi_f\left (0,1;\frac{a\tau+b}{c\tau+d}\right)=\Psi_f(c,d;\tau)\qquad \left(\begin{smallmatrix}
	a & b\\ c & d
\end{smallmatrix}\right)\in SL(2,\ZZ)\ ,
\ee so $\Psi_f(0,1;\tau)$ is invariant under 
\be \Gamma_0(2)=\left\{\left(\begin{smallmatrix}
	a & b\\ c & d
\end{smallmatrix}\right)\in SL(2,\ZZ)\mid c\equiv 0\mod 2\right\}\ .
\ee On the right-hand side of \eqref{eq:main2}, the $\tau$-dependent part $-f(0,1;\tau)$ is also $\Gamma_0(2)$-invariant. So both sides can be expanded as power series in $p$ with the coefficients being modular functions (more precisely, weakly holomorphic modular forms of weight $0$) for $\Gamma_0(2)$. The quotient $\CH/\Gamma_0(2)$ can be compactified by adjoining two points (cusps) corresponding to $\tau=0$ and $\tau=\infty$, and the resulting compact Riemann surface has genus zero. Thus, two weakly holomorphic modular forms of weight zero for $\Gamma_0(2)$ coincide if and only if they have the same polar and constant terms at both cusps. As a consequence, in order to prove the identity \eqref{eq:main2} it is sufficient to show that all polar and constant terms for $\tau\to \infty$ and $\tau\to 0$ are the same for both sides. For $\tau=\infty$, this means that  one needs to compare the expansions of both sides in powers of $q$ up to $q^0$. Since $C_{00}(n)=0$ for $n\le 0$, $C_{01}(n)=0$ for $n<0$, and $C_{01}(0)=-24$, for the left-hand side we get
\be p^{-1}\prod_{d=1}^\infty (1+p^d)^{-24}=\frac{\eta(\sigma)^{24}}{\eta(2\sigma)^{24}}\ .
\ee This corresponds to the first term on the right-hand side. Furthermore, the term 
$$24-f(0,1;\t)$$
vanishes up to $O(q^0)$, so we see that 
both sides  coincide at that order.

The expansion of \eqref{eq:main2} at $\tau\to 0$ is equivalent to the expansion of \eqref{eq:main1} at $\tau\to \infty$, since the two identities are related by $\tau\to -\frac{1}{\tau}$. By expanding the left-hand side of \eqref{eq:main1} up to $q^0$,  we get
\begin{align} p^{-1}(1-pq^{-1/2})&\prod_{d=1}^\infty (1-p^dq^{1/2})^{C_{d,1}(\frac{d}{2})}+O(q^{1/2})
\\&=p^{-1}-q^{-1/2}+\sum_{d=1}^\infty C_{d,1}(\frac{d}{2}) p^d+O(q^{1/2})\\
&=\sum_{n\in\ZZ} (C_{01}(\frac{n}{2})+C_{11}(\frac{n}{2}))p^n-q^{-1/2}+O(q^{1/2})\ ,
\end{align} where we used the fact that $C_{d,1}(n/2)=0$ unless $n\equiv d\mod 2$, and that $C_{11}(-1/2)=1$ is the only nonzero polar coefficient for $F_{01}$ and $F_{11}$.  One can see immediately that this expression equals the expansion 
\be f(1,0;2\sigma)-q^{-1/2}\ .
\ee of the right-hand side of \eqref{eq:main1} up to $q^0$. This concludes the proof.

\section{The simple roots}\label{sec:simple}
In the previous section we established modularity of the product side of the denominator formulas, one of which we had written down directly from the knowledge of the root system of our BKM. In this section we will verify that  the right hand side of equations (\ref{eq:main2}), (\ref{eq:main0})  can be interpreted as the additive side of a (super-)denominator identity, and at the same time we will determine the simple roots of the BKM superalgebra. 

Let us first rewrite the denominator in algebraic language; the product side, which appeared in \S \ref{sec:denom}, can be easily rewritten in this notation. For any super-BKM we can write its denominator formula as \cite{Ray}
\begin{equation}\label{eq:denom}
	e(-\rho)\sum_{w \in W}\det(w) w(T) = \dfrac{\prod_{\alpha \in \Delta_0^+}(1 - e(-\alpha))^{m_0(\alpha)}}{\prod_{\alpha \in \Delta_1^+}(1 + e(-\alpha))^{m_1(\alpha)}}.
\end{equation}
and the super-denominator as 
\begin{equation}\label{eq:superdenom}
	e(-\rho)\sum_{w \in W}\det(w) w(T') = \dfrac{\prod_{\alpha \in \Delta_0^+}(1 - e(-\alpha))^{m_0(\alpha)}}{\prod_{\alpha \in \Delta_1^+}(1 - e(-\alpha))^{m_1(\alpha)}}.
\end{equation}
In this formula, $\rho$ denotes the Weyl vector, $\Delta_{0, 1}^+$ denote the even and odd positive roots, respectively, $W$ denotes the Weyl group, $m_{0, 1}(\alpha)$ are the even and odd root multiplicities, respectively. $T$ is defined as follows. First, we define the heights of a root $\mu = \sum_{i \in I}k_i \alpha_i$ to be $ht(\mu) = \sum_{I}k_i, ht_0(\mu) = \sum_{I \backslash S} k_i$ where we use $I$ to denote the index set indexing the simple roots and $S \subseteq I$ indexes the odd generators only. Then we set $T = e(\rho)\sum  \epsilon(\mu)e(-\mu)$, with $\epsilon(\mu) = (-1)^{ht(\mu)}$. (One may similarly define a set $T'$ using $ht_0$). The sum is taken over all sums of imaginary simple roots, and $T$ is only nonvanishing if $\mu$ is the sum of pairwise orthogonal imaginary simple roots. Each simple root appears at least once. In addition, the Weyl group is generated by reflections with respect to the simple roots of positive norm. 

The BKM that appears in our story is of rank 2. It follows from Corollary 2.3.7 of \cite{Ray} that, in addition to the rank of the Cartan matrix, all of the structure in a given BKM is encoded in its set of roots, including their multiplicity, parity, and which among them are simple. In \S \ref{sec:mult} and \S \ref{sec:denom} we have deduced the multiplicity and parity of the root spaces that determine the product side of the denominator and superdenominator formulas. Thus, once we find the simple roots from the additive side of the (super)denominator formulas we will have completed our abstract characterization of the Conway BKM.

\subsection{Simple roots of the Conway Lie superalgebra}
We compute the additive side of the denominator identity for our BKM while concomitantly identifying the simple roots of the algebra. This information will also complete our abstract characterization of our BKM, by the remark at the end of the previous subsection.

Let us now find the additive side of the Weyl denominator identity conjectured in equation (\ref{eq:main2}). The corresponding \textit{super-}denominator formula is given by (\ref{eq:main0}). 

As explained in \S \ref{sec:algebra}, the roots are vectors in the even unimodular lattice $\Gamma^{1,1}$, and therefore can be labeled by a pair of integers $(d,r)\in \ZZ\oplus\ZZ$, with the norm of the root being $-2dr$. In particular, positive roots are of the form $(d, r), d>0$ with even multiplicities given by $C_{00}(rd)$ and odd multiplicities given by $|C_{01}(rd)|$. Note that for $rd \neq 0$, $C_{00}(rd) = - C_{01}(rd)$.  The result is that all terms in the product side of the super-denominator formula \eqref{eq:superdenom} cancel except those corresponding to $r=0$, where $C_{00}(0) =0, C_{10}(0)= -24$, i.e. there are 24 odd roots of length zero. In this case, the multiplicative and additive sides of the Weyl super-denominator formula are nothing but the product and sum forms of the inverse Dedekind-eta function. 

Let us show that the simple roots are all roots of the form $(1, r)$ or $(d, 0)$, $d>0$. To do this, we will first argue that such roots are necessarily simple. Next, in order to prove that this is a complete set of simple roots, we will check that they correctly reproduce the Weyl denominator and super-denominator. Notice that these are all imaginary simple roots of negative and null norm, respectively. Namely, roots of the form $(1, n), n<0$ would be real simple roots, but since both $F_{00}(\tau)$ and  $F_{01}(\tau)$ lack polar terms, there are no real simple roots of this form. 

The roots of type $(1, r)$ are clearly necessarily simple, as they cannot be expressed as the linear combination of other positive roots $(d,r)$, $d>0$, with nonnegative coefficients. There are $C_{00}(r)$ bosonic simple roots of this form and $|C_{01}(r)|$ fermionic simple roots. One can see that the null roots $(d, 0)$ are simple as follows. First notice that there are no positive roots $(d,r)$ with $r<0$, because $C_{00}(rd)=C_{01}(rd)=0$ for $rd<0$. Therefore, a root of the form $(d,0)$ can only be obtained as a sum of two positive roots $(d',0)$ and $(d'',0)$, with $d'+d''=d$. There are no even roots of the form $(d, 0)$ with $d>0$, because $C_{00}(0)=0$, so any two algebra elements $u\in \g(d', 0)$ and $v\in \g(d'', 0)$ are necessarily odd. But then their anticommutator $[u,v]$ must be an even element in $\g_0(d'+d'',0)$, and, plainly, there are no even roots of this form. Thus, any two elements $u\in \g(d',0)$, $v\in \g(d'',0)$ must anticommute with one another. Since this implies that one cannot get a root of the form $(d,0)$ via commutators or anticommutators of any other positive roots, they are simple.   

Now, assuming the the roots $(1,r)$ and $(d,0)$ are the only simple roots, let us compute the additive side of the denominator and super-denominator formula, and check that it matches with the product side. With this assumption on the simple roots, the Weyl vector is given by $\rho=(-1,0)$. Due to the absence of real simple roots, the Weyl group must be trivial. Therefore, the additive side of the denominator formula is simply equal to $T$, and that of the superdenominator to $T'$. Let us focus on the denominator formula, which is of greater interest to us, and tally first the simple roots of the form $(1, r)$: \be T \supset p^{-1}(1 - p\sum_{r>0} C_{00}(r) q^r + p \sum_{r \geq 0}C_{01}(r) q^r) = p^{-1} - F_{00}(\t) + F_{01}(\t)\ \ee

Next, we need to tally the contributions from the roots of the form $(d, 0)$. Recall that they each contribute with multiplicity 24. Then
\bea\nonumber T &\supset& {1 \over p} \sum_{m=1}^{\infty} p^m \sum_{\lambda \vdash m} (-1)^{l(\lambda)}\prod_{j}\begin{pmatrix} 23 + \lambda_j \\ \nonumber \lambda_j \end{pmatrix} \\ & =& (1/p)\prod_{n=1}^{\infty}(1 + p^{n})^{-24}- p^{-1} = \left({\eta(\sigma) \over \eta(2\sigma)}\right)^{24} - p^{-1}.\eea In the first expression, we are summing over all partitions $\lambda$ of $m$, $l(\lambda)$ denotes the number of integers in the partition, $j$ indexes the integers in the partition and $\lambda_j$ denotes the multiplicity of $j$. Indeed, the first two expressions are just the number of ways to write a simple root $(m, 0)$ as sums of $l(\lambda)$ roots $(k, 0), k \leq m$, with each root contributing a minus sign. 

In total, on the additive side we have $F_{01}(\t)- F_{00}(\t) + ({\eta(\sigma) \over \eta(2\sigma)})^{24} + 24 = ({\eta(\sigma) \over \eta(2\sigma)})^{24}+24 - f(0,1;\t)$ (the $+24$ correction is needed not to overcount the roots $(1,0)$, that were included among both the $(d,0)$ and the $(1,r)$ roots). This is the same as the right-hand side of eq.\eqref{eq:main2}, so it perfectly matches the product formula $p^{-1}\Psi_f(0,1;\tau)$ for the denominator. 

The additive side of the superdenominator is simply obtained by flipping the signs of the contributions from the odd simple roots. Therefore, roots of the form $(1, r)$ contribute
\be p^{-1} - F_{00}(\t) - F_{01}(\t)=p^{-1}+24
\ee and the roots of the form $(d,0)$ contribute
\be (1/p)\prod_{n=1}^{\infty}(1 - p^{n})^{-24}- p^{-1} =\frac{1}{\eta(\sigma)^{24}}-p^{-1}\ ,
\ee so that the additive side of the super-denominator formula is $\frac{1}{\eta(\sigma)^{24}}$ (again, taking into account the $-24$ correction for the $(1,0)$ root), which coincides with the right-hand side of \eqref{eq:main0} (since $-24-f(0,0;\tau)=0$) and therefore correctly reproduces the product formula. We conclude that our assumption about the simple roots is correct.

The identification of the simple roots of $\g$ allows us to determine the homology spaces $H_i(\g_+)$ and write a homological version of the denominator formula. For any algebra $\g$, $H_0(\g_+)$ is always one dimensional and $H_1(\g_+)$ is spanned by the simple roots, so that
\bea \nonumber &H_0(\g_+)=\CC\ ,\\ \nonumber
&H_1(\g_+)=\bigoplus_{d>0}p^d\g(d,0)\oplus \bigoplus_{r>0}pq^r\g(1,r)\ .\nonumber 
\eea 
In general, the space $H_i(\g_+)$ is spanned by the elements of $\bigwedge\nolimits^i\g_+$ whose degree $k$ satisfies \be\label{homolcond} (k+\rho)^2=\rho^2\ ,\ee where $\rho$ is the Weyl vector \cite{BorcherdsMM,GarlandLepowsky}. In our case, $\rho=(-1,0)$, so $\rho^2=0$, so for an element of degree $(d,r)$  condition \eqref{homolcond} translates into
\be r(d-1)=0\ .
\ee The degrees of elements in $\bigwedge\nolimits^i\g_+$ satisfy $d\ge i$, so for $i>1$ this condition can only be satisfied if $r=0$. It follows that
\be H_i(\g_+)=\bigwedge\nolimits^i \left (\bigoplus_{d>0}p^d\g(d,0)\right),\qquad i\ge 2\ .
\ee It follows that
\be H(\g_+)= \bigwedge \left (\bigoplus_{d>0}p^d\g(d,0)\right )\oplus \bigoplus_{r>0}pq^r\g(1,r)\ ,
\ee which as a $Co_0$ module is isomorphic to \footnote{Here, the twisted module $V_{tw}^{s\natural}$, and in particular its component $V_{tw}^{s\natural}(0)$, are regarded as superspaces, with the $\ZZ_2$-grading given by the fermion number, so that the exterior product $u\wedge v$ is $\ZZ_2$-graded skew-symmetric. In particular, $V_{tw}^{s\natural}(0)$ has trivial even component, so any two elements $u,v\in V_{tw}^{s\natural}(0)$ are necessarily odd and their exterior product satisfies  $u\wedge v=v\wedge u$.}
\be\label{addCo0module} \bigwedge \left (\bigoplus_{d>0}p^dV_{tw}^{s\natural}(0)\right)\oplus \bigoplus_{r>0}pq^rV_{tw}^{s\natural}(r)\ .
\ee 
 Notice that, while for the Monster Lie algebra $H_i=0$ for $i>2$, in this case the $H_i(\g_+)$ are  all non-trivial, due to the presence of roots of zero norm. 

The explicit construction of the homology spaces shows that $\bigwedge \g_+=H(\g_+)$ is an identity between virtual, $\Gamma^{1,1}$- and $\ZZ_2$-graded, $Co_0$ modules. This is the starting point for the derivation of the twisted denominator and super-denominator identities in section \ref{sec:twisted}.

\subsection{Simple roots for the other denominator}
It is tempting to conjecture that equation (\ref{eq:main1}) is also the super-denominator identity for a BKM algebra\footnote{This denominator identity and the denominator for the Conway Lie algebra should be thought of as expansions of a single formula at different cusps in the domain of definition (physically: arising from a single string compactification where the cusps represent perturbative duality frames which are identified by the action of the duality group). The  expansions at the two cusps correspond to different BKMs, each of which captures the perturbative BPS spectrum in the corresponding duality frame; from this point of view the states appearing in the expansion around the other cusp are non-perturbative. We do not construct the algebra whose denominator formula produces this identity, and it would be fascinating to give a physical interpretation of the relationship between the two cusps.} Assuming that this is true, one can use the super-denominator formula to derive the multiplicities of the positive roots and the simple roots.  On the product side we have
\begin{equation}
	\prod_{d=1}^{\infty}\prod_{r \in \ZZ}(1 -  p^d q^{r/2})^{C_{d, r}(rd/2)}.
\end{equation}
Since $C_{0,0}(n),C_{1,1}(n)\ge 0$ and $C_{0,1}(n),C_{1,0}(n)\le 0$, we conjecture that the positive roots of degree $(d,r)$ have multiplicity $|C_{d,r}(rd/2)|$, and they are even or odd depending on whether $d+r$ is even or odd, respectively.  The Weyl group is generated by reflections with respect to the real simple roots. In this case, the only real roots  are $\pm(1, -1)$, and they are necessarily simple. The contribution of the positive root to the product formula comes from the exponent  $C_{1,1}(-1/2)$, i.e. when $r=-1, d=1$. Because we only have a single real root up to sign, the Weyl group is just $\ZZ/2\ZZ$.

Given this, let us try to reconstruct the additive side of the super-denominator formula. We conjecture that the only simple roots are of the form $(1, r)$, and therefore are odd if $r$ is even and are even if $r$ is odd. Notice that the multiplicity of roots with $r=0$ is $|C_{d,0}(0)|=0$, and hence no null roots are present. The rest of the computation proceeds quite analogously to the case of the Monster Lie algebra \cite{BorcherdsMM}.  The Weyl vector is $\rho=(-1,0)$. Computing the inner product of two simple roots shows that no imaginary simple roots are orthogonal to one another. Hence, $T'$ computed in one Weyl chamber is just the sum of simple roots: 
\be T'= p^{-1}(1 - p\sum_{r>0} C_{1,1}(r/2) q^{r/2} - p \sum_{r>0} C_{1,0}(r/2)q^{r/2}) = p^{-1} +  q^{-1/2} - f(1,0;\t)\ ,\ee
where we used the fact that the multiplicities of the even and the odd roots of degree $(1,r)$ are $C_{1,1}(r/2)$ and  $-C_{1,0}(r/2)$, respectively.

Next, we Weyl reflect the result through the unique simple root of positive norm, $(1, -1)$ to obtain $w(T')$. The Weyl reflection should take $(1, 0)$ to $(0, 1)$, which translates to $q \rightarrow p^2$. Then $r(T') = q^{-1/2} + p^{-1} - f(1,0;2\sigma)$. 

Putting the pieces of the additive side together we have $e(-\rho)(T' - r(T')) = p(p^{-1} + q^{-1/2} - f(1,0;\t) - p^{-1} - q^{-1/2} + f(1,0;2\sigma)) = p(f(1,0;2\sigma) - f(1,0;\tau))$, as desired. 

A similar computation may be done for the denominator formula. Given our assumptions about the positive and the simple roots of the algebra, the product side becomes
\begin{equation}
\prod_{d=1}^{\infty}\prod_{r \in \ZZ}(1 - (-1)^{d+r} p^d q^{r/2})^{C_{d, r}(rd/2)}\ ,
\end{equation}
while on the additive side
\be T= p^{-1}(1 - p\sum_{r>0} C_{1,1}(r/2) q^{r/2} + p \sum_{r>0} C_{1,0}(r/2)q^{r/2}) = p^{-1} +  q^{-1/2} - f(1,1;\t)\ ,\ee
so that $e(-\rho)(T - r(T)) = p(p^{-1} + q^{-1/2} - f(1,1;\t) - p^{-1} - q^{-1/2} + f(1,1;2\sigma)) = p(f(1,1;2\sigma) - f(1,1;\tau))$. So, if our assumptions are correct, one should have an identity
\be \prod_{d=1}^{\infty}\prod_{r \in \ZZ}(1 - (-1)^{d+r} p^d q^{r/2})^{C_{d, r}(rd/2)}=p(f(1,1;2\sigma) - f(1,1;\tau))\ .
\ee
 But this follows immediately from identity \eqref{eq:main3}, by replacing $p\to -p$ (equivalently, $\sigma\to \sigma+\frac{1}{2}$) on both sides, and noting that $f(1,1;2\sigma+1)=f(1,0;2\sigma)$.

\section{Conway action and twisted denominators}\label{sec:twisted}
Now that we have the denominator identities for our Conway Lie algebra firmly in hand, we compute their appropriate twisted counterparts. As described at the end of \S \ref{sec:denom}, these follow from the homological formulation of the Weyl denominator identity plus the $Co_0$ action on the Conway Lie algebra $\g$. 
\subsection{Twisted denominators of the Conway BKM superalgebra}\label{sec:conwaytwist}
In this section we will define a pair of twisted denominator and superdenominator formulas for our BKM for each element $g \in Co_0$. 

Let us define a set of $g$-twined functions for $g\in Co_0$ as,
\bea
f_g(1,1;\t)&=& -\Tr_{V^{s\natural}} g q^{L_0-c/24}\\
f_g(1,0;\t)&=& \sTr_{V^{s\natural}} g q^{L_0-c/24}\\
f_g(0,1;\t)&=& \Tr_{V^{s\natural}_{tw}} g q^{L_0-c/24}\\
f_g(0,0;\t)&=& \sTr_{V^{s\natural}_{tw}} g q^{L_0-c/24}\ .
\eea 
Note that these functions only depend on the conjugacy class $[g]\in Co_0$. One can define these functions explicitly in terms of the Frame shape of $g$, $\pi_g$ \cite{Duncan:2014eha}. The Frame shape $\pi_g$ of an element $g \in Co_0$ of order $N$ is a symbol of the form $\prod_{\ell|N}\ell^{m_\ell}$, for some $m_\ell\in \ZZ$, that encodes the information about the eigenvalues of $g$ in the standard $24$-dimensional representation $\rho_{\bf 24}$ of $Co_0$. In particular, when $g$ acts by a permutation in this representation, $\pi_g$ is just the cycle shape. In general, the integers $m_\ell$ are determined by
\be \det(t-\rho_{\bf 24}(g))=\prod_{\ell|N}(t^\ell-1)^{m_\ell}\ .
\ee
We have that
\bea
f_g(1,1;\t)&=& -{1\over 2} \left ({\eta_g(\t/2)\over \eta_g(\t)} + {\eta_{-g}(\t/2)\over \eta_{-g}(\t)}- {\cal C}_g \eta_g(\t) + {\cal C}_{-g} \eta_{-g}(\t)\right )\\
f_g(1,0;\t)&=& {1\over 2} \left ({\eta_g(\t/2)\over \eta_g(\t)} + {\eta_{-g}(\t/2)\over \eta_{-g}(\t)}+ {\cal C}_g \eta_g(\t) - {\cal C}_{-g} \eta_{-g}(\t)\right )\\
f_g(0,1;\t)&=& {1\over 2} \left (-{\eta_g(\t/2)\over \eta_g(\t)} + {\eta_{-g}(\t/2)\over \eta_{-g}(\t)}+ {\cal C}_g \eta_g(\t) + {\cal C}_{-g} \eta_{-g}(\t)\right )\\
f_g(0,0;\t)&=& {1\over 2} \left ({\eta_g(\t/2)\over \eta_g(\t)} - {\eta_{-g}(\t/2)\over \eta_{-g}(\t)}+ {\cal C}_g \eta_g(\t) + {\cal C}_{-g} \eta_{-g}(\t)\right ),
\eea
where one can find the explicit definitions in \cite{Duncan:2014eha}. Here, for an element of Frame shape $\pi_g=\prod_{\ell|N}\ell^{m_\ell}$, the eta product $\eta_g$ is defined as
\be \eta_g(\tau)=\prod_{\ell|N}\eta(\ell\tau)^{m_\ell}\ .
\ee Furthermore, $-g\in Co_0$ denotes the product of $g$ by the generator of the centre $\ZZ_2\subset Co_0$. The coefficients ${\cal C}_g$ are determined up to a sign by the eigenvalues of $g$ and are tabulated in \cite{Duncan:2014eha}.  Briefly, one writes the $24$ eigenvalues of $g$ in $\rho_{\textbf{24}}$ as $e^{\pm 2i\alpha_k}$, for suitable $\alpha_1,\ldots,\alpha_{12}\in 2\pi\mathbb{Q}$, and sets ${\cal C}_g=\prod_{k=1}^{12}(e^{i\alpha_k}-e^{-i\alpha_k})$.

We will let the coefficients of the functions $f_g(i,j;\t)$ be given by
\be\label{eq:gcoeff}
f_g(i,j;\t)=\sum_n c^g_{i,j}(n/2) q^{n/2}.
\ee
Finally, we introduce the functions 
\be
F_{r,s}^g(\t) =\sum_n C_{r,s}^g(n) q^n={1\over 2}\left (f_g(r,0;\t) + (-1)^sf_g(r,1;\t)\right )
\ee
which capture the spectrum of states invariant and anti-invariant under $(-1)^F$ in each sector, such that
\be
C^g_{r,s}(n)= {1\over 2}(c^g_{r,0}(n) + (-1)^s c^g_{r,1}(n)).
\ee

Let $N=o(g)$ and define 
\be
\hat c^g_{r,s}(n,k):= {1\over N}\sum_{j=0}^{N-1} e^{-{2\pi i jk\over N}}c^{g^j}_{r,s}(n)
\ee
where the $c^g$ are defined in (\ref{eq:gcoeff}), and similarly
\be
\hat C^g_{r,s}(n,k):= {1\over 2}(\hat c^g_{r,0}(n,k) + (-1)^s \hat c^g_{r,1}(n,k)).
\ee
From these coefficients, define the products
\bea\nn
p^{-1} \Psi_{f_g}(0,1;\t)&:=& p^{-1} \prod_{d=1}^\infty \prod_{r\in \mathbb Z} \prod_{k=0}^{N-1}(1-e^{2\pi i k/N}p^d q^{r})^{\hat C^g_{0,0}({rd},k)}(1+e^{2\pi i k/N}p^d q^{r})^{\hat C^g_{0,1}({rd},k)}\\\nn
p^{-1} \Psi_{f_g}(0,0;\t)&:=& p^{-1} \prod_{d=1}^\infty \prod_{r\in \mathbb Z} \prod_{k=0}^{N-1}(1-e^{2\pi i k/N}p^d q^{r})^{\hat C^g_{0,0}({rd},k)+\hat C^g_{0,1}({rd},k)}\eea These products correspond, respectively, to the $g$-twisted denominator and super-denominator for the BKM superalgebra $\g$, i.e. to the trace and supertrace of $g$ over the $\Gamma^{1,1}$- and $\ZZ_2$-graded $Co_0$-module $\bigwedge \g_+$. Indeed, from the explicit description of $\bigwedge \g_+$ as a $Co_0$-module given in the previous section, it follows that $\hat C^g_{0,0}(rd,k)$ and $|\hat C^g_{0,1}(rd,k)|$ are the multiplicities of, respectively, even and odd positive roots of degree $(d,r)$, that are  eigenvectors of $g$ with eigenvalue $e^{\frac{2\pi i k}{N}}$. Each such root contributes a factor $(1\mp e^{2\pi i k/N}p^d q^{r})^{\pm 1}$ to the trace and $(1- e^{2\pi i k/N}p^d q^{r})^{\pm 1}$ to the supertrace.  By the isomorphism $\bigwedge \g_+\cong H(\g_+)$ of $Co_0$-modules, these products must equal the corresponding trace and super-trace of $g$ over the homology $H(\g_+)$. This construction produces a $g$-twisted (super-)denominator identity of the form:
\bea\label{Tw00}
p^{-1} \Psi_{f_g}(0,0;\t)&=&{1\over \eta_g(\sigma)}-\chi_g - f_g(0,0;\t)\\\label{Tw01}
p^{-1} \Psi_{f_g}(0,1;\t)&=&{1 \over \eta_{-g}(\sigma)}-\chi_{-g}-f_g(0,1;\t)
\eea where, on the additive side, $\chi_g = \Tr_{\bf 24}g$,   ${1\over \eta_g(\sigma)}$ (respectively, ${1 \over \eta_{-g}(\sigma)}$) is the contribution\footnote{In order to compute the contribution from $\bigwedge (\bigoplus_{d>0}p^dV_{tw}^{s\natural}(0))$, it is important to remember that $V_{tw}^{s\natural}(0)$ is a superspace with trivial even part, so the wedge product $u_1\wedge \ldots\wedge u_n$ of any elements $u_1,\ldots,u_n\in V_{tw}^{s\natural}(0)$ is completely \emph{symmetric} under permutations.} from $\bigwedge (\bigoplus_{d>0}p^dV_{tw}^{s\natural}(0))$ in \eqref{addCo0module}, while  $-\chi_g - f_g(0,0;\t)$ (respectively, $-\chi_{-g}-f_g(0,1;\t)$) comes from $\bigoplus_{r>0}pq^rV_{tw}^{s\natural}(r)$. 
\subsection{Twisted denominators at the other cusp}
Under the assumption that equation \eqref{eq:main2} is the superdenominator of a BKM super-algebra with an action of $Co_0$, we can define the following products corresponding to the $g$-twisted denominators of this algebra
\bea\nn
p^{-1} \Psi_{f_g}(1,0;\t)&:=& p^{-1} \prod_{d=1}^\infty \prod_{r\in \mathbb Z} \prod_{k=0}^{N-1}(1-e^{2\pi i k/N}p^d q^{r\over 2})^{\hat C^g_{d,r}({rd\over 2},k)}\\\nn
p^{-1} \Psi_{f_g}(1,1;\t)&:=& p^{-1} \prod_{d=1}^\infty \prod_{r\in \mathbb Z} \prod_{k=0}^{N-1}(1-e^{2\pi i k/N}(-1)^r p^d q^{r\over 2})^{\hat C^g_{d,r}({rd\over 2},k)}.
\eea
These products produce twisted denominator identities of the form:
\bea\label{Tw10}
p^{-1} \Psi_{f_g}(1,0;\t)&=&f_g(1,0; 2\sigma)-f_g(1,0;\t)\\\label{Tw11}
p^{-1} \Psi_{f_g}(1,1;\t)&=&f_g(1,0; 2\sigma)-f_g(1,1;\t).
\eea
Though we have not constructed a BKM algebra corresponding to these denominators, we checked the validity of several of these twisted identities. This gives strong evidence that such an algebra exists and  carries a natural action of the group $Co_0$. We hope to return to this question in the future.
\subsection{Examples}
In this section we explicitly verify the identities (\ref{Tw00}), (\ref{Tw01}), (\ref{Tw10}) and (\ref{Tw11}) for all elements $g \in Co_0$ of order 2 and order 3. 

\subsubsection{Order 2}
For $N=2$, we have that $\hat C^g_{r,s}(n,0)= {1\over 2}(C^e_{r,s}(n) + C^g_{r,s}(n))$ and $\hat C^g_{r,s}(n,1)= {1\over 2}(C^e_{r,s}(n) - C^g_{r,s}(n)).$ There are four conjugacy classes in $Co_0$ of order 2; we present the necessary data and twisted identities here. For each such conjugacy class $g$ we specify below the Frame shape $\pi_g$, dual Frame shape, $\pi_{-g}$, and the data $\{{\cal C}_g, {\cal C}_{-g}\}$ derived from the Frame shapes and described in \S \ref{sec:conwaytwist}. Finally we present the first few terms in the product expansion of the corresponding twisted identities.

\begin{itemize}
\item $g=2A; \pi_g=2^{24}/1^{24}, \pi_{-g}=1^{24}; {\cal C}_g=4096, {\cal C}_{-g}=0$.
\bea\nn
F_{00}^{2A}(\t)&=&{1\over 2}(f_{2A}(0,0;\t)+f_{2A}(0,1;\t))=2048 q + 49152 q^2 + \ldots\\\nn
F_{01}^{2A}(\t)&=&{1\over 2}(f_{2A}(0,0;\t)-f_{2A}(0,1;\t))=24+2048 q + 49152 q^2  + \ldots\\\nn
F_{10}^{2A}(\t)&=&{1\over 2}(f_{2A}(1,0;\t)+f_{2A}(1,1;\t))=2048 q + 49152 q^2 + \ldots\\\nn
F_{11}^{2A}(\t)&=&{1\over 2}(f_{2A}(1,0;\t)-f_{2A}(1,1;\t))=q^{-1/2} + 276q^{1/2} + 11202 q^{3/2}+ \ldots
\eea
From our twisted denominator and superdenominator we find the following twisted identities:
\bea\nn
p^{-1}\Psi_{f_{2A}}(0,0;\t)&=&{1\over p}\prod_{d=1}^\infty {1\over (1+p^d)^{24}} {(1-pq)^{2048}\over (1+pq)^{2048}}{(1-pq^2)^{49152}\over (1+pq^2)^{49152}}{(1-p^2q)^{49152}\over (1+p^2q)^{49152}}\cdots\\\nn
&=&{\eta^{24}(\sigma)\over \eta^{24}(2\sigma)}+24-f_{2A}(0,0;\t)\\\nn
p^{-1}\Psi_{f_{2A}}(0,1;\t)&=&{1\over p}\prod_{d=1}^\infty {1\over (1-p^d)^{24}}={1\over \eta^{24}(\sigma)}-24-f_{2A}(0,1;\t).
\eea

Similarly, at the other cusp we find the following identities: 
\bea\nn
p^{-1}\Psi_{f_{2A}}(1,0;\t)&=&{1\over p} {(1-pq^{-{1\over 2}}) (1-pq^{1\over2})^{276}(1-p^3q^{1\over 2})^{11202} (1-pq^{3\over2})^{11202}(1-p^2q)^{49152}\cdots \over (1+p^2q^{1\over 2})^{2048}(1+pq)^{2048} (1+p^4q^{1\over 2})^{49152}\cdots}\\\nn
&=&f_{2A}(1,0; 2\sigma)-f_{2A}(1,0;\t)\\\nn
p^{-1}\Psi_{f_{2A}}(1,1;\t)&=&{1\over p} {(1+pq^{-{1\over 2}}) (1+pq^{1\over 2})^{276}(1+p^3q^{1\over 2})^{11202} (1+pq^{3\over 2})^{11202}(1-p^2q)^{49152}\cdots \over (1-p^2q^{1\over 2})^{2048}(1+pq)^{2048} (1-p^4q^{1\over 2})^{49152}\cdots}\\\nn
&=&f_{2A}(1,0; 2\sigma)-f_{2A}(1,1;\t).
\eea
\item $g=2B; \pi_g = 1^82^8, \pi_{-g} = {2^{16}\over1^8}; {\cal C}_{\pm g}=0$.
\bea\nn
F_{00}^{2B}(\t)&=&{1\over 2}(f_{2B}(0,0;\t)+f_{2B}(0,1;\t))=0\\\nn
F_{01}^{2B}(\t)&=&{1\over 2}(f_{2B}(0,0;\t)-f_{2B}(0,1;\t))=-8\\\nn
F_{10}^{2B}(\t)&=&{1\over 2}(f_{2B}(1,0;\t)+f_{2B}(1,1;\t))=0\\\nn
F_{11}^{2B}(\t)&=&{1\over 2}(f_{2B}(1,0;\t)-f_{2B}(1,1;\t))=q^{-1/2} + 20q^{1/2} -62 q^{3/2} + \ldots
\eea
where $F_{11}^{2B}(\t)=\eta(\t/2)^8/\eta(2\t)^8+8$.

The twisted identities for our algebra are 
\bea\nn
p^{-1}\Psi_{f_{2B}}(0,0;\t)&=&{1\over p}\prod_{d=1}^\infty {1\over (1-p^d)^{16}(1+p^d)^8} ={1\over\eta^8(\sigma) \eta^{8}(2\sigma)}-8-f_{2B}(0,0;\t)\\\nn
p^{-1}\Psi_{f_{2B}}(0,1;\t)&=&{1\over p}\prod_{d=1}^\infty {1\over (1+p^d)^{16}(1-p^d)^8}={\eta^8(\sigma)\over \eta^{16}(2\sigma)}+8-f_{2B}(0,1;\t),
\eea
and at the other cusp we find the following identities
\bea\nn
p^{-1}\Psi_{f_{2B}}(1,0;\t)&=& {(1-pq^{-{1\over 2}}) (1-pq^{1\over2})^{148}(1+pq^{1\over2})^{128}(1-p^3q^{1\over 2})^{5570}(1+p^3q^{1\over 2})^{5632}\cdots \over p(1-p^2q^{1\over 2})^{1024}(1+p^2q^{1\over 2})^{1024}(1-pq)^{1024}(1+pq)^{1024} \cdots}\\\nn
&=&f_{2B}(1,0; 2\sigma)-f_{2B}(1,0;\t)\\\nn
p^{-1}\Psi_{f_{2B}}(1,1;\t)&=&{(1+pq^{-{1\over 2}}) (1+pq^{1\over2})^{148}(1-pq^{1\over2})^{128}(1+p^3q^{1\over 2})^{5570}(1-p^3q^{1\over 2})^{5632}\cdots \over p(1-p^2q^{1\over 2})^{1024}(1+p^2q^{1\over 2})^{1024}(1-pq)^{1024}(1+pq)^{1024} \cdots}\\\nn
&=&f_{2B}(1,0; 2\sigma)-f_{2B}(1,1;\t).
\eea

\item $g=2C;  \pi_{g} = {2^{16}\over1^8},\pi_{-g} = 1^82^8; {\cal C}_{\pm g}=0$.
\bea\nn
F_{00}^{2C}(\t)&=&{1\over 2}(f_{2C}(0,0;\t)+f_{2C}(0,1;\t))=0\\\nn
F_{01}^{2C}(\t)&=&{1\over 2}(f_{2C}(0,0;\t)-f_{2C}(0,1;\t))=+8\\\nn
F_{10}^{2C}(\t)&=&{1\over 2}(f_{2C}(1,0;\t)+f_{2C}(1,1;\t))=0\\\nn
F_{11}^{2C}(\t)&=&{1\over 2}(f_{2C}(1,0;\t)-f_{2C}(1,1;\t))=q^{-1/2} + 20q^{1/2} -62 q^{3/2} + \ldots
\eea

In this case, the twisted identities for our algebra are 
\bea\nn
p^{-1}\Psi_{f_{2C}}(0,0;\t)&=&{1\over p}\prod_{d=1}^\infty {1\over (1+p^d)^{16}(1-p^d)^8} ={\eta^8(\sigma)\over \eta^{16}(2\sigma)}+8-f_{2C}(0,0;\t)\\\nn
p^{-1}\Psi_{f_{2C}}(0,1;\t)&=&{1\over p}\prod_{d=1}^\infty {1\over (1-p^d)^{16}(1+p^d)^8}={1\over\eta^8(\sigma) \eta^{8}(2\sigma)}-8-f_{2C}(0,1;\t).
\eea
Note that here we find $\Psi_{f_{2B}}(0,0;\t)=\Psi_{f_{2C}}(0,1;\t)$ and $\Psi_{f_{2B}}(0,1;\t)=\Psi_{f_{2C}}(0,0;\t)$.

At the other cusp we find the following identities
\bea\nn
p^{-1}\Psi_{f_{2C}}(1,0;\t)&=& {(1-pq^{-{1\over 2}}) (1-pq^{1\over2})^{148}(1+pq^{1\over2})^{128}(1-p^3q^{1\over 2})^{5570}(1+p^3q^{1\over 2})^{5632}\cdots \over p(1-p^2q^{1\over 2})^{1024}(1+p^2q^{1\over 2})^{1024}(1-pq)^{1024}(1+pq)^{1024} \cdots}\\\nn
&=&f_{2C}(1,0; 2\sigma)-f_{2C}(1,0;\t)\\\nn
p^{-1}\Psi_{f_{2C}}(1,1;\t)&=&{(1+pq^{-{1\over 2}}) (1+pq^{1\over2})^{148}(1-pq^{1\over2})^{128}(1+p^3q^{1\over 2})^{5570}(1-p^3q^{1\over 2})^{5632}\cdots \over p(1-p^2q^{1\over 2})^{1024}(1+p^2q^{1\over 2})^{1024}(1-pq)^{1024}(1+pq)^{1024} \cdots}\\\nn
&=&f_{2C}(1,0; 2\sigma)-f_{2C}(1,1;\t).
\eea
Note that we find $\Psi_{f_{2B}}(1,0;\t)=\Psi_{f_{2C}}(1,0;\t)$ and $\Psi_{f_{2B}}(1,1;\t)=\Psi_{f_{2C}}(1,1;\t)$.

\item $g=2D; \pi_g = 2^{12}, \pi_{-g} = 2^{12}; {\cal C}_{\pm g}=0$.
\bea\nn
F_{00}^{2D}(\t)&=&{1\over 2}(f_{2D}(0,0;\t)+f_{2D}(0,1;\t))=0\\\nn
F_{01}^{2D}(\t)&=&{1\over 2}(f_{2D}(0,0;\t)-f_{2D}(0,1;\t))=0\\\nn
F_{10}^{2D}(\t)&=&{1\over 2}(f_{2D}(1,0;\t)+f_{2D}(1,1;\t))=0\\\nn
F_{11}^{2D}(\t)&=&{1\over 2}(f_{2D}(1,0;\t)-f_{2D}(1,1;\t))=q^{-1/2} -12 q^{1/2} +66q^{3/2} + \ldots
\eea
where $F_{11}^{2D}(\t)=\eta(\t)^{12}/\eta(2\t)^{12}$.

In this case the the twisted identities for are algebra are
\bea\nn
p^{-1}\Psi_{f_{2D}}(0,0;\t)&=&{1\over p}\prod_{d=1}^\infty {1\over (1-p^d)^{12}(1+p^d)^{12}} ={1\over\eta^{12}(2\sigma)}-0-f_{2C}(0,0;\t)\\\nn
p^{-1}\Psi_{f_{2D}}(0,1;\t)&=&{1\over p}\prod_{d=1}^\infty {1\over (1+p^d)^{12}(1-p^d)^{12}}={1\over\eta^{12}(2\sigma)}+0-f_{2C}(0,1;\t),
\eea
i.e., we have that $\Psi_{f_{2D}}(0,0;\t)=\Psi_{f_{2D}}(0,1;\t)$, whereas  at the other cusp we find the following identities
\bea\nn
p^{-1}\Psi_{f_{2D}}(1,0;\t)&=& {(1-pq^{-{1\over 2}}) (1-pq^{1\over2})^{132}(1+pq^{1\over2})^{144}(1-p^3q^{1\over 2})^{5634}(1+p^3q^{1\over 2})^{5568}\cdots \over p(1-p^2q^{1\over 2})^{1024}(1+p^2q^{1\over 2})^{1024}(1-pq)^{1024}(1+pq)^{1024} \cdots}\\\nn
&=&f_{2D}(1,0; 2\sigma)-f_{2D}(1,0;\t)\\\nn
p^{-1}\Psi_{f_{2D}}(1,1;\t)&=&{(1+pq^{-{1\over 2}}) (1+pq^{1\over2})^{132}(1-pq^{1\over2})^{144}(1+p^3q^{1\over 2})^{5634}(1-p^3q^{1\over 2})^{5568}\cdots \over p(1-p^2q^{1\over 2})^{1024}(1+p^2q^{1\over 2})^{1024}(1-pq)^{1024}(1+pq)^{1024} \cdots}\\\nn
&=&f_{2D}(1,0; 2\sigma)-f_{2D}(1,1;\t).
\eea

\subsubsection{Order 3}
For $N=3$, we have that $\hat C^g_{r,s}(n,0)= {1\over 3}(C^e_{r,s}(n) +2 C^g_{r,s}(n))$ and $\hat C^g_{r,s}(n,1)=\hat C^g_{r,s}(n,2)= {1\over 3}(C^e_{r,s}(n) - C^g_{r,s}(n)).$ There are four conjugacy classes in $Co_0$ of order 3. For each such conjugacy class $g$ we specify below the Frame shape $\pi_g$, dual Frame shape, $\pi_{-g}$, and the data $\{C_g, C_{-g}\}$ derived from the Frame shapes and described in \S \ref{sec:conwaytwist}. Finally we present the first few terms in the product expansion of the corresponding twisted identities.
\item $g=3A; \pi_g=3^{12}/1^{12}, \pi_{-g}=1^{12}6^{12}/2^{12}3^{12}; {\cal C}_g=729, {\cal C}_{-g}=1$. 
\bea\nn
F_{00}^{3A}(\t)&=&{1\over 2}(f_{3A}(0,0;\t)+f_{3A}(0,1;\t))=365q + 4368q^2+32838q^3+\ldots \\\nn
F_{01}^{3A}(\t)&=&{1\over 2}(f_{3A}(0,0;\t)-f_{3A}(0,1;\t))=12 + 364 q + 4380 q^2 +32772 q^3+ \ldots\\\nn
F_{10}^{3A}(\t)&=&{1\over 2}(f_{3A}(1,0;\t)+f_{3A}(1,1;\t))=364 q + 4380 q^2 +32772 q^3+ \ldots\\\nn
F_{11}^{3A}(\t)&=&{1\over 2}(f_{3A}(1,0;\t)-f_{3A}(1,1;\t))=q^{-1/2}+78 q^{1/2}+ 1365 q^{3/2}+ \ldots
\eea
Let us denote the primitive third roots of unity by
\be \omega:=e^{\frac{2\pi i}{3}}\ ,\qquad \bar\omega:=e^{-\frac{2\pi i}{3}}\ .
\ee
From our twisted denominator and superdenominator we find the following twisted identities:
\bea\nn
p^{-1}\Psi_{f_{3A}}(0,0;\t)&=&{1\over p}\prod_{d=1}^\infty {1\over (1-\omega p^d)^{12}(1-\bar\omega p^d)^{12} }{(1-pq)^{486}(1-pq^2)^{5832}(1-p^2q)^{5832}\cdots \over (1-\omega pq)^{243}(1-\bar\omega pq)^{243}\cdots }\\\nn
&=&{\eta^{12}(\sigma)\over \eta^{12}(3\sigma)}+12-f_{3A}(0,0;\t)\\\nn
p^{-1}\Psi_{f_{3A}}(0,1;\t)&=&{1\over p}\prod_{d=1}^\infty {1\over (1+\omega p^d)^{12}(1+\bar\omega p^d)^{12} }{(1-pq)^{926}(1-\omega pq)^{561}(1-\bar\omega pq)^{561}\cdots \over (1+pq)^{440}(1+\omega pq)^{804}(1+\bar\omega pq)^{804}\cdots }\\\nn&=&{\eta^{12}(2\sigma)\eta^{12}(3\sigma)\over \eta^{12}(\sigma)\eta^{12}(6\sigma)}-12-f_{3A}(0,1;\t).
\eea
Similarly, at the other cusp we find the following identities: 
\bea\nn
p^{-1}\Psi_{f_{3A}}(1,0;\t)&=&{(1-pq^{-{1\over 2}}) (1-pq^{1\over2})^{144}(1-\omega pq^{1\over2})^{66}(1-\bar\omega pq^{1\over2})^{66}\cdots \over p(1-p^2q^{1\over 2})^{440}(1-\omega p^2q^{1\over2})^{804}(1-\bar\omega p^2q^{1\over2})^{804} \cdots}\\\nn
&=&f_{3A}(1,0; 2\sigma)-f_{3A}(1,0;\t)\\\nn
p^{-1}\Psi_{f_{3A}}(1,1;\t)&=&{(1+pq^{-{1\over 2}}) (1+pq^{1\over2})^{144}(1+\omega pq^{1\over2})^{66}(1+\bar\omega pq^{1\over2})^{66}\cdots \over p(1+p^2q^{1\over 2})^{440}(1+\omega p^2q^{1\over2})^{804}(1+\bar\omega p^2q^{1\over2})^{804} \cdots}\\\nn
&=&f_{3A}(1,0; 2\sigma)-f_{3A}(1,1;\t).
\eea

\item $g=3B; \pi_g = 1^63^6, \pi_{-g} = {2^{6}6^6\over1^63^6}; {\cal C}_{g}=0, {\cal C}_{-g}=64$.
\bea\nn
F_{00}^{3B}(\t)&=&{1\over 2}(f_{3B}(0,0;\t)+f_{3B}(0,1;\t))=32 q + 192 q^2 + 672 q^3 + \ldots \\\nn
F_{01}^{3B}(\t)&=&{1\over 2}(f_{3B}(0,0;\t)-f_{3B}(0,1;\t))=-6-32 q -192 q^2 - 672 q^3 + \ldots\\\nn
F_{10}^{3B}(\t)&=&{1\over 2}(f_{3B}(1,0;\t)+f_{3B}(1,1;\t))=-32 q -192 q^2 - 672 q^3 + \ldots\\\nn
F_{11}^{3B}(\t)&=&{1\over 2}(f_{3B}(1,0;\t)-f_{3B}(1,1;\t))=q^{-1/2} + 15q^{1/2} +87 q^{3/2} + \ldots
\eea
From our twisted denominator and superdenominator we find the following twisted identities:
\bea\nn
p^{-1}\Psi_{f_{3B}}(0,0;\t)&=&{1\over p}\prod_{d=1}^\infty {1\over (1-p^d)^{12}(1-\omega p^d)^{6}(1-\bar\omega p^d)^{6} }\\\nn
&=&{1\over \eta^6(\sigma)\eta^{6}(3\sigma)}-6-f_{3B}(0,0;\t)\\\nn
p^{-1}\Psi_{f_{3B}}(0,1;\t)&=&{1\over p}\prod_{d=1}^\infty {1\over (1+p^d)^{12}(1+\omega p^d)^{6}(1+\bar\omega p^d)^{6} }{(1-pq)^{704}(1-\omega pq)^{672}(1-\bar\omega pq)^{672}\cdots \over (1+pq)^{704}(1+\omega pq)^{672}(1+\bar\omega pq)^{672}\cdots }\\\nn
&=&{\eta^{6}(\sigma)\eta^{6}(3\sigma)\over \eta^{6}(2\sigma)\eta^{6}(6\sigma)}+6-f_{3B}(0,1;\t).
\eea
Similarly, at the other cusp we find the following identities: 
\bea\nn
p^{-1}\Psi_{f_{3B}}(1,0;\t)&=&{(1-pq^{-{1\over 2}}) (1-pq^{1\over2})^{102}(1-\omega pq^{1\over2})^{87}(1-\bar\omega pq^{1\over2})^{87}\cdots \over p(1-p^2q^{1\over 2})^{704}(1-\omega p^2q^{1\over2})^{672}(1-\bar\omega p^2q^{1\over2})^{672} \cdots}\\\nn
&=&f_{3B}(1,0; 2\sigma)-f_{3B}(1,0;\t)\\\nn
p^{-1}\Psi_{f_{3B}}(1,1;\t)&=&{(1+pq^{-{1\over 2}}) (1+pq^{1\over2})^{102}(1+\omega pq^{1\over2})^{87}(1+\bar\omega pq^{1\over2})^{87}\cdots \over p(1+p^2q^{1\over 2})^{704}(1+\omega p^2q^{1\over2})^{672}(1+\bar\omega p^2q^{1\over2})^{672} \cdots}\\\nn
&=&f_{3B}(1,0; 2\sigma)-f_{3B}(1,1;\t).
\eea

\item $g=3C;  \pi_{g} = {3^9/1^3},\pi_{-g} = 1^36^9/2^33^9; {\cal C}_g=0, {\cal C}_{-g}=-8$.
\bea\nn
F_{00}^{3C}(\t)&=&{1\over 2}(f_{3C}(0,0;\t)+f_{3C}(0,1;\t))=-4 q + 12 q^2-12 q^3+\ldots\\\nn
F_{01}^{3C}(\t)&=&{1\over 2}(f_{3C}(0,0;\t)-f_{3C}(0,1;\t))=3+4 q - 12 q^2+12 q^3+\ldots\\\nn
F_{10}^{3C}(\t)&=&{1\over 2}(f_{3C}(1,0;\t)+f_{3C}(1,1;\t))=4 q - 12 q^2+12 q^3+\ldots\\\nn
F_{11}^{3C}(\t)&=&{1\over 2}(f_{3C}(1,0;\t)-f_{3C}(1,1;\t))=q^{-1/2} + 6q^{1/2} -3 q^{3/2} + \ldots
\eea
From our twisted denominator and superdenominator we find the following twisted identities:
\bea\nn
p^{-1}\Psi_{f_{3C}}(0,0;\t)&=&{1\over p}\prod_{d=1}^\infty {1\over (1-p^d)^{6}(1-\omega p^d)^{9}(1-\bar\omega p^d)^{9} }\\\nn
&=&{\eta^3(\sigma)\over \eta^{9}(3\sigma)}+3-f_{3C}(0,0;\t)\\\nn
p^{-1}\Psi_{f_{3C}}(0,1;\t)&=&{1\over p}\prod_{d=1}^\infty {1\over (1+p^d)^{6}(1+\omega p^d)^{9}(1+\bar\omega p^d)^{9} }{(1-pq)^{680}(1-\omega pq)^{684}(1-\bar\omega pq)^{684}\cdots \over (1+pq)^{680}(1+\omega pq)^{684}(1+\bar\omega pq)^{684}\cdots }\\\nn
&=&{\eta^{3}(2\sigma)\eta^{9}(3\sigma)\over \eta^{3}(\sigma)\eta^{9}(6\sigma)}-3-f_{3C}(0,1;\t).
\eea
Similarly, at the other cusp we find the following identities: 
\bea\nn
p^{-1}\Psi_{f_{3C}}(1,0;\t)&=&{(1-pq^{-{1\over 2}}) (1-pq^{1\over2})^{96}(1-\omega pq^{1\over2})^{90}(1-\bar\omega pq^{1\over2})^{90}\cdots \over p(1-p^2q^{1\over 2})^{680}(1-\omega p^2q^{1\over2})^{684}(1-\bar\omega p^2q^{1\over2})^{684} \cdots}\\\nn
&=&f_{3C}(1,0; 2\sigma)-f_{3C}(1,0;\t)\\\nn
p^{-1}\Psi_{f_{3C}}(1,1;\t)&=&{(1+pq^{-{1\over 2}}) (1+pq^{1\over2})^{96}(1+\omega pq^{1\over2})^{90}(1+\bar\omega pq^{1\over2})^{90}\cdots \over p(1+p^2q^{1\over 2})^{680}(1+\omega p^2q^{1\over2})^{684}(1+\bar\omega p^2q^{1\over2})^{684} \cdots}\\\nn
&=&f_{3C}(1,0; 2\sigma)-f_{3C}(1,1;\t).
\eea

\item $g=3D; \pi_g = 3^{8}, \pi_{-g} = 6^8/3^8; {\cal C}_g=0, {\cal C}_{-g}=16$.
\bea\nn
F_{00}^{3D}(\t)&=&{1\over 2}(f_{3D}(0,0;\t)+f_{3D}(0,1;\t))=8q + 64 q^4 + \ldots\\\nn
F_{01}^{3D}(\t)&=&{1\over 2}(f_{3D}(0,0;\t)-f_{3D}(0,1;\t))=-8q - 64 q^4 + \ldots\\\nn
F_{10}^{3D}(\t)&=&{1\over 2}(f_{3D}(1,0;\t)+f_{3D}(1,1;\t))=-8q - 64 q^4 + \ldots\\\nn
F_{11}^{3D}(\t)&=&{1\over 2}(f_{3D}(1,0;\t)-f_{3D}(1,1;\t))=q^{-1/2} +28q^{5/2} + \ldots
\eea
From our twisted denominator and superdenominator we find the following twisted identities:
\bea\nn
p^{-1}\Psi_{f_{3D}}(0,0;\t)&=&{1\over p}\prod_{d=1}^\infty {1\over (1-p^d)^{8}(1-\omega p^d)^{8}(1-\bar\omega p^d)^{8} }\\\nn
&=&{1\over \eta^{8}(3\sigma)}-f_{3D}(0,0;\t)\\\nn
p^{-1}\Psi_{f_{3D}}(0,1;\t)&=&{1\over p}\prod_{d=1}^\infty {1\over (1+p^d)^{8}(1+\omega p^d)^{8}(1+\bar\omega p^d)^{8} }{(1-pq)^{688}(1-\omega pq)^{680}(1-\bar\omega pq)^{680}\cdots \over (1+pq)^{688}(1+\omega pq)^{680}(1+\bar\omega pq)^{680}\cdots }\\\nn
&=&{\eta^{8}(3\sigma)\over \eta^{8}(6\sigma)}-f_{3D}(0,1;\t).
\eea
Similarly, at the other cusp we find the following identities: 
\bea\nn
p^{-1}\Psi_{f_{3D}}(1,0;\t)&=&{(1-pq^{-{1\over 2}}) (1-pq^{1\over2})^{92}(1-\omega pq^{1\over2})^{92}(1-\bar\omega pq^{1\over2})^{92}\cdots \over p(1-p^2q^{1\over 2})^{688}(1-\omega p^2q^{1\over2})^{680}(1-\bar\omega p^2q^{1\over2})^{680} \cdots}\\\nn
&=&f_{3D}(1,0; 2\sigma)-f_{3D}(1,0;\t)\\\nn
p^{-1}\Psi_{f_{3D}}(1,1;\t)&=&{(1+pq^{-{1\over 2}}) (1+pq^{1\over2})^{92}(1+\omega pq^{1\over2})^{92}(1+\bar\omega pq^{1\over2})^{92}\cdots \over p(1+p^2q^{1\over 2})^{688}(1+\omega p^2q^{1\over2})^{680}(1+\bar\omega p^2q^{1\over2})^{680} \cdots}\\\nn
&=&f_{3D}(1,0; 2\sigma)-f_{3D}(1,1;\t).
\eea
\end{itemize}

\section{Conclusion}\label{sec:conc}

In this note, we have constructed a Borcherds Kac-Moody algebra with natural action of the Conway sporadic group $Co_0$ and proved associated (super)denominator identities using equivariant Hecke operators. This Conway Lie algebra may be obtained by applying a certain functor to the super vertex operator algebra (SVOA) of Duncan, analogously to parallel constructions in Monstrous moonshine \cite{BorcherdsMM, Carnahan, Carnahan2}. We apply this technology at the level of the physics of a (chiral) superstring worldsheet, after Borcherds and Scheithauer, using BRST cohomology and the no-ghost theorem. The resulting denominator identities may provide an alternative route to proving the genus zero property of the McKay-Thompson series of the Conway module, following the logic of \cite{PPV, PPV2}, i.e. without recourse to the Hauptmoduls of Monstrous moonshine. We have outlined a family of compactifications in type IIA string theory where this program may be carried out in full. Physically, we anticipate that this model will provide another concrete instantiation of the conjecture by Harvey and Moore \cite{HM1, HM2} that Borcherds Kac-Moody algebras govern algebras of BPS states in certain string vacua; in particular, we expect that the Conway Lie algebra will play the role of a spectrum generating algebra over which the spacetime BPS states form a module. In this context, the (twisted) denominator formulas should appear as supersymmetric indices. 

We emphasize that a different BKM algebra (the ``fake Monster Lie algebra'')  with a natural action of an extension $\ZZ_2^{24}.Co_0$ of the Conway group was already proposed by Borcherds \cite{BorcherdsFake}, and its twisted denominators identities were analyzed in \cite{Sch3,Sch4}. It should be possible to obtain this algebra as a ``BPS algebra'' in a compactification of the heterotic string, building on the worldsheet theory discussed in \cite{phi12}, and by repeating the construction in \cite{PPV,PPV2} with the Monster VOA $V^\natural$ with the lattice VOA based on the Leech lattice.   It would be interesting to understand whether there are any connections (maybe through string dualities) with the algebra in the present paper.

The denominator $\Psi_f(0,1;\t) $ of the BKM superalgebra is the expansion at one cusp of an automorphic form on $\CH\times \CH$, invariant under a subgroup $\Gamma\subset O(2,2,\ZZ)$. As suggested in \S \ref{sec:proof} and \ref{sec:simple},  the expansion of this automorphic form at a different cusp of $(\CH\times \CH)/\Gamma$ provides the infinite products $\Psi_f(1,0;\t) $ and $\Psi_f(1,1;\t) $. Conjecturally, these products correspond to the denominator and super-denominator of a different BKM superalgebra with an action of $Co_0$, and therefore give rise to new twisted denominator identities. The expansions at the two cusps are related by a $O(2,2,\ZZ)$ transformation acting on the automorphic form. In a full (not chiral) superstring theory framework, as the one sketched in \S \ref{sec:strings}, one expects this automorphic group to admit an interpretation in terms of dualities. Hence, we expect that a detailed understanding of the approach outlined in \S \ref{sec:strings} will lead to a uniform construction of both algebras, and to a proof of our conjectures.

One central, outstanding question in moonshine is: what is the most appropriate characterization of the (mock) modular forms appearing as graded traces of moonshine modules? Reformulations and extensions of the genus zero property, notably Rademacher summability \cite{DF, CDRad, CDRad2, SarahFrancesca, CDHJacobi},
have received much attention; for the vector-valued mock modular forms of umbral moonshine related characterizations based on analyticity of the modular forms (e.g. ``optimality'') \cite{CDH1, CDH2, CDOpt} have been proposed. On the other hand, the construction of \cite{PPV, PPV2} relates growth properties of the McKay-Thompson series at cusps to the behavior of supersymmetric indices under natural physical operations such as dualities and decompactification. Perhaps this philosophy---in addition to providing a physical derivation of the genus zero property of Conway moonshine---may give clues towards taming the zoo of modular objects in moonshine. Of course, for Monstrous and Conway moonshines, we have explicit moonshine modules in hand. Notably, a uniform construction of the umbral moonshine modules is still lacking (see, however, \cite{ACH, CDMod, DH, DOD}).  

Another outstanding question in moonshine is that of the connection between umbral moonshine (particularly $M_{24}$ moonshine) and the geometry of K3 surfaces. In $M_{24}$ moonshine, the mock modular form appears in a decomposition of the K3 elliptic genus, a Jacobi form \cite{EOT}. Analogous decompositions of the K3 elliptic genus related to all instances of umbral moonshine were proposed in \cite{CH}. In the context of K3 surfaces, the Conway module serves as a device that generates putative twining genera---the Jacobi form analogue of McKay-Thompson series---associated to symmetries of SCFTs on the moduli space of nonlinear sigma models with K3 target \cite{DuncanMC}. Such symmetry groups are in one-to-one  correspondence with the so-called 4-plane preserving subgroups of Co$_0$; these are the subgroups that preserve at least a 4-dimensional subspace in its natural geometric action on the Leech lattice \cite{GHV}\footnote{The reader is advised to consult \cite{CHVZ} for precise conjectures and some proofs on the relationship between symmetries and twining genera in K3 SCFTs and the Jacobi forms motivated from moonshine \cite{CH, DuncanMC}.}. In this application, one promotes the Conway module to an $\CN=4$ SVOA as described in \cite{M5}, whose chiral algebra includes in particular an affine $SU(2)$ subalgebra (R-symmetry) which does not commute with the supercurrents. This promotion furnishes the twining genera with an elliptic variable, conjugate to the $U(1)$ charge in the Cartan of $SU(2)$, that is part and parcel of a Jacobi form. Multiplicative lifts of all twining genera in \cite{DuncanMC}, which produce Siegel forms encapsulating BPS state counts in string theory on (orbifolds of) $K3 \times T^2$, have been produced \cite{PVZ}. It would be extremely interesting to promote our construction of the Conway Lie algebra, such that our denominators acquire an extra variable, and clarify its connection to algebras of these spacetime BPS states in string theory. We hope to return to this point in the future.

As a final point, we note that the multiplicative side of the (twisted) denominator identities of our Conway Lie algebra admit an alternative derivation. The functions may be obtained via the Borcherds, or theta, lift, by applying Theorem 13.3 of \cite{Borcherds:1996uda}. 
To build a Borcherds product based on a generic lattice $M$, one needs to consider a vector-valued modular form $F$ whose components $F_\gamma$ are in one to one correspondence with the cosets $\gamma\in M^\vee/M$. This vector valued modular form should transform in the representation $\rho_M$ that is given by
\be F_\gamma(\tau+1)=e(\gamma^2/2) F_\gamma(\tau)
\ee
\be F_{\gamma}(-1/\tau)=\frac{i^{(b^+-b^-)/2}}{\sqrt{|M^\vee/M|}}\sum_{\delta\in M^\vee/M}e(-(\gamma,\delta))F_\delta\ ,
\ee where $(b^+,b^-)$ is the signature of $M$.
In our case, the lattice $M$ is $U\oplus U(2)$, with $M^\vee=U\oplus \frac{1}{2}U(2)$ and $M^\vee/M\cong \frac{1}{2}U(2)/U(2)\cong \ZZ_2\times \ZZ_2$. In this notation $U$ is the standard hyperbolic lattice and $U(n)$ is the same lattice with its intersection form rescaled by $n$. The elements of $\frac{1}{2}U(2)$ can be written as $\lambda=(\frac{r}{2},\frac{s}{2})$, with $\lambda^2=rs$ and the cosets of $M^\vee/M\cong \frac{1}{2}U(2)/U(2)$ are labeled by pairs $r,s\mod 2$. Therefore, the representation $\rho_M$ is spanned by four functions $F_{r,s}$, $m,n\in \ZZ/2\ZZ$, that should satisfy
\be F_{r,s}(\tau+1)=(-1)^{rs} F_{r,s}(\tau)
\ee
\be F_{r,s}(-1/\tau)=\frac{1}{2}\sum_{r',s'\in \ZZ/2\ZZ} (-1)^{rs'+r's} F_{r',s'}\ .
\ee
These are the $F_{r, s}(\tau)$ that appear in the main text. The only complication is to verify that at each of the two cusps the Weyl vector $e(\rho, Z)$ is $1/p$ and that the Weyl chamber condition $(\lambda, W)>0$ restricts the product over $d, r \in \ZZ$ to $d>0$; we leave this step for the interested reader to complete.

\bigskip

\noindent{\bf Acknowledgements.}
We thank S. Carnahan, J. Duncan, D. Persson, N. Scheithauer and T. Wrase for useful comments and discussions. We are especially grateful to J. Duncan for helpful comments on an earlier version of the draft. S.H. is supported by the National Science and Engineering Council of Canada and the Canada Research Chairs program. N.P. is supported by a Sherman Fairchild Postdoctoral Fellowship. R.V. is supported by a grant from `Programma per giovani ricercatori Rita Levi Montalcini'. The authors acknowledge support from the Aspen Center for Physics, which is supported by National Science Foundation grant PHY-1607611. N.P. gratefully acknowledges the Perimeter Institute of Theoretical Physics for hospitality and support during the final stages of this work. This material is based upon work supported by the U.S. Department of Energy, Office of Science, Office of High Energy Physics, under Award Number DE-SC0011632.
\appendix

\section{Borcherds Kac-Moody superalgebras}\label{a:BKM}

In this appendix we discuss the definition and some of the main properties of Borcherds-Kac-Moody (BKM)  superalgebras\footnote{We will work with BKMs over $\CC$ that have nontrivial odd part.} that will be useful in the main text, following \cite{Ray}. The reader may also wish to consult \cite{borcherds1988generalized, borcherds1995automorphic, Borcherds:1996uda, Sch1, Sch2, jurisich1996exposition, jurisich1998generalized}. We then verify that the Lie superalgebra constructed in the main text is a BKM.

BKM algebras are a generalization of Kac-Moody Lie algebras. Recall that the roots of a Lie algebra $L$ with Cartan subalgebra $H$ are the elements $\alpha\in H^*$ such that,  $[h, l] = \alpha(h) l$ for some $l\in L$, and for all $ h \in H$, and the simple roots are the generating set. Notably, BKMs frequently have imaginary (simple) roots, that is, roots $\alpha$ of nonpositive norm, $(\alpha, \alpha) \leq 0$. Many properties of BKMs are quite similar to those of their Kac-Moody counterparts and much of the highest weight theory, for instance, may be applied with only minor modifications; see for instance the Weyl-Kac formula for BKMs in Section \ref{sec:simple}. For instance, a BKM $L$ admits the usual triangular decomposition $L= L_+\oplus H \oplus L_-$ and one defines a Weyl group of reflections with respect to the positive roots of nonzero norm.\footnote{The full Weyl group has a subgroup called the even Weyl group generated by reflections $r_i$ for $i \in I$ such that $a_{ii}>0$.}

Each BKM $L$ will be associated to a Cartan matrix $A$ and an abelian Lie algebra $H$. The input data is characterized as follows. Take $H_{\RR}$ a real vector space with elements $h_i, i \in I$, where $I$ is some (finite or countably infinite) indexing set. In addition, the vector space must be equipped with a symmetric, non-degenerate, real-valued bilinear form $( , )$.

More precisely, the bilinear form on a Lie superalgebra $L$ (i.e. a $\ZZ_2$-graded Lie algebra equipped with a Lie bracket satisfying the usual properties) is chosen to satisfy (Definition 2.2.2, \cite{Ray}):
\begin{enumerate}
\item Supersymmetry: A bilinear form $(,)$ is said to be supersymmetric if $(x, y) = (-1)^{deg(x)deg(y)}(y, x)$ for $x, y \in L$. \\
\item Consistency: A bilinear form $(,)$ is said to be consistent if $(L_{even}, L_{odd}) = 0$.\\
\item Invariance: A bilinear form $(,)$ is said to be invariant if $([g, x], y) = (g, [x, y])$ for all elements $g, x, y \in L$.
\end{enumerate} Here $deg(v)$ denotes the degree with respect to the $\ZZ_2$ (i.e. even or odd) grading.

The inner products of the elements $h_i$ with respect to this bilinear form satisfy  conditions: $(h_i, h_j) \leq 0$ if $i \neq j$; if $(h_i, h_i) > 0$, then $2 (h_i, h_j)/(h_i, h_i) \in \ZZ$ for all $j \in I$; if $(h_i, h_i) >0$ and $i \in S$ (with $S \subseteq I$ the indexing set of odd generators), then $(h_i, h_j) \in \ZZ$ for all $j \in I$. Take $H = H_{\RR}\otimes_{\RR}\CC$. The Cartan matrix $A$ is given in terms of its matrix entries $a_{ij}:= (h_i, h_j)$. When working with finite Lie superalgebras, one takes the usual Killing form $K(x, y):= \text{str}(\text{ad}(x) \text{ad}(y))$ to be $(,)$. 

The bilinear form is defined on $H$ but extends to all of $G$.

Then we define a BKM $L$ to be a Lie superalgebra (equipped with $(,)$ as above) generated by elements $h_i \in H$ and Chevalley-Serre generators $e_i, f_i, i \in I$,  satisfying the defining relations:
\begin{enumerate}
\item $[e_i, f_j] = \delta_{ij} h_j$ \\
\item $[h, e_i] = (h, h_i)e_i, \ [h, f_i] = -(h, h_i)f_i $ \\
\item $deg(e_i) = 0 = deg(f_i)$ if $i \notin S$; $deg(e_i) = 1 = deg(f_i)$ if $i \in S$\\
\item $(ad(e_i))^{1-2 a_{ij}/a_{ii}}e_j = 0 = (ad(f_i))^{1-2 a_{ij}/a_{ii}}f_j$ if $a_{ii}>0$ and $i \neq j$\\
\item $(ad(e_i))^{1- a_{ij}/a_{ii}}e_j = 0 = (ad(f_i))^{1- a_{ij}/a_{ii}}f_j$ if $i \in S, a_{ii}>0$ and $i \neq j$\\
\item $[e_i, e_j] = 0 = [f_i, f_j]$ if $a_{ij}=0$.
\end{enumerate} Notice that if $a_{ii}>0$ for all $i \in I$ these are the defining relations of a Kac-Moody superalgebra.

In principle, one can check that a Lie superalgebra is a BKM by checking these defining relations. In practice, however, verifying that a Lie superalgebra is a BKM is more easily done by verifying certain properties of the nondegenerate bilinear form. In particular, BKMs may be characterized among Lie superalgebras as follows:
\begin{theorem} (Theorem 2.5.4, \cite{Ray})
	A Lie superalgebra $L$ that satisfies the following conditions is a BKM superalgebra:
	\begin{enumerate}
		\item There is a self-centralizing even subalgebra $H$ such that $L$ is a direct sum of eigenspaces of $H$. Each eigenspace is finite dimensional. A root of $L$ is  a non-zero eigenvalue of $H$.
		\item There is a non-degenerate invariant supersymmetric bilinear form $(\cdot,\cdot)$ on $L$.
		\item There is an element $h\in H$ (called a \emph{regular element}) such that the subalgebra of $L$ commuting with $h$ is $H$ (i.e., there are no roots orthogonal to $h$), and such that for all $r>0$ there are only finitely many roots $\alpha$ with $0<|\alpha(h)|<r$. We call a root $\alpha$ positive if $\alpha(h)>0$ and negative if $\alpha(h)<0$.
		\item The norms of the roots are bounded from above.
		\item Let $\alpha$ and $\beta$ be both positive or both negative roots of non-positive norm (imaginary roots). Then, $(\alpha,\beta)\le 0$. Moreover, if $(\alpha,\beta)=0$ and if $x\in L_\alpha$ and $[x,L_{-\gamma}]=0$ for all roots $\gamma$ with $0<\gamma(h)< |\alpha(h)|$, then $[x,L_\beta]=0$
	\end{enumerate}
\end{theorem}
Let us use this characterization to demonstrate that our Lie superalgebra is a BKM. 

For the algebra we are interested in, $H\equiv \g(0)$ is two-dimensional with signature $(1,1)$ with respect to the bilinear form, and the roots span an even unimodular lattice $\Gamma^{1,1}\subset H^*$. The elements of this lattice can be written as $\alpha :=\left\lbrace r,d \right\rbrace \in \ZZ\oplus\ZZ$, with norm $(\alpha, \alpha)=-2rd$. The eigenspaces of $H$ are the components $\g(k)$, $k\in \Gamma^{1,1}$, which are finite dimensional. All roots are imaginary (i.e., with zero or negative norms), so their norms are automatically bounded from above. A regular element $h\in H$ can be taken of the form $\left\lbrace R,1/R \right\rbrace\subset \RR^{1,1}$, provided that $R$ is irrational. 

To check condition $5$, we notice that for an algebra graded in a space of signature $(1,1)$, two positive (or two negative) roots $\alpha,\beta$ of non-positive norm necessarily satisfy $(\alpha,\beta)\le 0$, and they satisfy $(\alpha,\beta)=0$ if and only if they have both null norm and they are proportional to each other. For the algebra we are considering, the roots of zero norm only correspond to odd generators, so that $[L_\alpha,L_\beta]$ is a subspace of the even part of $L_{\alpha+\beta}$. On the other hand, $L_{\alpha+\beta}$ has only odd generators, because $\alpha+\beta$ has zero norm, so $[L_\alpha,L_\beta]=0$ and the last condition is automatically satisfied.
\section{Construction of the algebra: further details}\label{a:algebradetails}

It is useful to consider the bosonizations of the fermion space-time fermions $\psi^\mu$, the ghosts $b,c$ and the superghosts $\beta,\gamma$. All these fields can be described in terms of a lattice vertex algebra for four bosons $\lambda,\sigma,\chi,\phi$ on the lattice   
\be L^{\lambda,\phi}\oplus L^{\chi,\sigma}\ ,\ee
where
\be L^{\lambda ,\phi}=\{(x_1,x_2)\in \RR^{1,1}\mid \text{all }x_i\in \ZZ\text{ or all }x_i\in \frac{1}{2}+\ZZ\}\ ,\ee
and
\be L^{\chi,\sigma}=\ZZ^2\ .\ee
More precisely, the $bc$ system is described as
\be\label{ghosts} b=e^{-\sigma}\qquad c=e^{\sigma}\ ,
\ee while the matter space-time fermions and the superghosts in the NS sector (corresponding to vectors $(x_1,x_2)\in L^{\lambda,\phi}$ with $x_1,x_2\in \ZZ$) are given by
\be \psi^{\pm}\sim e^{\pm\lambda}\ ,\qquad :\psi^+\psi^-:\sim \partial\lambda\ ,
\ee
and
\be\label{superghosts} \beta=\partial \chi e^{-\phi+\chi}\qquad \gamma=e^{\phi-\chi}\ .
\ee  In the Ramond sector (corresponding vectors $(x_1,x_2)\in L^{\lambda,\phi}$ with $x_1,x_2\in \frac{1}{2}+\ZZ$), the spin fields $\Theta_\alpha$ of weight $5/8$, where $\alpha=\pm$ represent the two chiralities of a  spinor in $1+1$ space-time dimensions, are given by
\be \Theta_{\pm}=e^{\pm\frac{1}{2}\lambda}
\ee and they always come combined with some exponential $e^{(n+\frac{1}{2})\phi}$, $n\in \ZZ$. In particular, one has the weight $1/2$ fields
\be S_{\pm}=\Theta_{\pm}e^{-\phi/2}=e^{\pm\frac{1}{2}\lambda-\phi/2}\ .
\ee 

Specifically, $L^{\lambda,\phi}$ contains the even sublattice $L^{\lambda,\phi}_0=\{(x_1,x_2)\in \RR^{1,1}\mid \text{all }x_i\in \ZZ\text{ and }\sum_i x_i\in 2\ZZ\}$ and the quotient $L^{\psi,\phi}/L^{\psi,\phi}_0\cong \ZZ_2\times \ZZ_2$ has four cosets denoted by $A,V,S,C$. Also the ghost lattice $L^{\chi,\sigma}$ contains an even sublattice $L^{\chi,\sigma}_0$ spanned by elements of even norm, and $L^{\chi,\sigma}/L^{\chi,\sigma}_0\cong \ZZ_2$, so that the different sectors of the theory can be labeled by pairs of cosets in $L^{\psi,\phi}/L^{\psi,\phi}_0$ and $L^{\chi,\sigma}/L^{\chi,\sigma}_0$.

After this bosonization, the $V_{1,1}$ matter SCFT, the ghost and the superghost systems can all be described in terms of a lattice vertex algebra on the lattice
\be L^X\oplus L^{\psi,\phi}\oplus L^{\chi,\sigma}\ ,\ee
where, \be L^X\cong \Gamma^{1,1}\ee  is isomorphic to the even unimodular lattice of signature $(1,1)$. The corresponding vertex algebra contains the bosonic fields in the 1+1 space-time directions $\partial X^\mu$, $\mu=0,1$ and $e^{ikX}$.  The VA related to the lattice $L^X\oplus L^{\psi,\phi}\oplus L^{\chi,\sigma}$ has $8$ different sectors labeled by the four cosets $A,V,S,C$ in $L^{\psi,\phi}/L^{\psi,\phi}_0\cong \ZZ_2\times \ZZ_2$ and the two cosets $0,1$ in $L^{\chi,\sigma}/L^{\chi,\sigma}_0\cong \ZZ_2$.   The full GSO projected space $V_{GSO}$ is obtained by tensoring in a suitable way these $8$ cosets with the four sectors of internal SVOA $V^{f\natural}$, namely $V^{f\natural}_{NS+},V^{f\natural}_{NS-},V^{f\natural}_{R+},V^{f\natural}_{R-}$:
\be \CH_{GSO}  =\oplus_{i=0}^1 [(V_{NS+}\otimes  V_{(A,i)})\oplus (V_{NS-}\otimes  V_{(V,i)})\oplus (V_{R+}\otimes  V_{(S,i)})\oplus (V_{R-}\otimes  V_{(C,i)})]\ .
\ee  

\section{Zero momentum cohomology}\label{a:zeromomentum}
In this section, we provide a description of the spaces $H(k)_{-1,1}$ and  $H(k)_{-1/2,1}$ for $k^2=0$. The derivation is completely analogous to the one in \cite{Sch1}, though the final results are different. 

Let us start with the $-1$ picture. The spaces $C_{-1,n}(k)$ for $k^2=0$ and $n=0,1$ (considered in Proposition 5.3 of \cite{Sch1}) are:
	\begin{itemize}
		\item $C(k)_{-1,1}$ is two dimensional and spanned  by $\psi^\mu_{-1/2}e^{-\phi}_{-1/2}c_{1}|k\rangle$, $\mu=0,1$. These states only involve the ground state in the internal VOA $V_{NS}\cong V^{f\natural}$. This is different than in \cite{Sch1}, since in that case the internal VOA contains some fields $\upsilon$ of weight $1/2$ in the NS sector, so that $C(k)_{-1,1}$ would contain additional states of the form $\upsilon_{-1/2}e^{-\phi}_{-1/2}c_{1}|k\rangle$.
		\item $C(k)_{-1,0}$ is spanned by $e^{-2\phi}_{0}(\partial\xi)_{-1}c_{1}|k\rangle$.
	 \end{itemize}
The cohomology groups $H(k)_{-1,1}$, considered in Proposition 5.11 of \cite{Sch1}, in our case are given by
	\begin{itemize}
		\item for $k^2=0$, $k\neq 0$, one has $H(k)_{-1,1}=0$ (one state in $C(k)_{-1,1}$ is eliminated by the $Q$-closedness condition, and the remaining one is $Q$-exact);
		\item for $k=0$, $H(k)_{-1,1}$ is spanned by the two states $\psi^\mu_{-1/2}e^{-\phi}_{-1/2}c_{1}|0\rangle$, $\mu=0,1$ (the momentum generators in $-1$ picture).
	\end{itemize} 
To summarize, in the $-1$ picture, for $k^2=0$, we have
\be
\begin{matrix}
	& \dim H(k)_{-1,1} \\
	k_0=\pm k_1,\ k\neq 0	& 0 \\
	k=0 & 2	
\end{matrix}
\ee
The two generators of $H(0)_{-1,1}$ correspond to the space time momenta $P^\mu$, $\mu=0,1$.

Let us now consider the half-integral picture numbers. The spaces $C_{p,n}(k)$, $p=-1/2,-3/2$, $n=0,1$, for $k^2=0$  (considered in Proposition 5.3 of \cite{Sch1}) are:
	\begin{itemize}
		\item $C(k)_{-1/2,1}$ is spanned by the $24$ states $(\chi^i_-\Theta_+e^{-\phi/2})_{-1}c_1|k\rangle$, where $\chi_-^i$, $i=1,\ldots, 24$, is one of the $24$ ground states in $V_{R-}\subset V_R\cong V^{f\natural}_{tw}$ and $\Theta_+=e^{\frac{\lambda}{2}}$ is a spin field corresponding to a vector of conformal weight $1/8$ in the lattice vertex algebra $L^{\lambda,\phi}$. For a general internal VOA, such as the one in \cite{Sch1}, one has also the states $(\chi^i_+\Theta_-e^{-\phi/2})_{-1}c_1|k\rangle$, where $\chi^i_+$ are states of weight $1/2$ in the $R+$ sector.
		\item The space $C(k)_{-1/2,0}$ is zero.
		\item The space $C(k)_{-3/2,1}$ is $24$ dimensional and a basis is given by  $(\chi^i_-\Theta_-e^{-3\phi/2})_{-1}c_1|k\rangle$ (for a general VOA, there are also states $(\chi^i_+\Theta_+e^{-3\phi/2})_{-1}c_1|k\rangle$)
		\item The space $C(k)_{-3/2,0}$ is $24$ dimensional and a basis is given by  $(\chi_-\Theta_+e^{-5\phi/2})_{0}(\partial\xi)_{-1}c_1|k\rangle$ (in a general VOA, also the states $(\chi_+\Theta_-e^{-5\phi/2})_{0}(\partial\xi)_{-1}c_1|k\rangle$) \end{itemize}
 The cohomology groups $H(k)_{-1/2,1}$, considered in Proposition 5.11 of \cite{Sch1}, in our case are given by
	\begin{itemize}
		\item for $k^0=-  k^1$, $k\neq 0$, one has $H(k)_{-1/2,1}=0$. Indeed, $Q$-closedness translates into the massless Dirac equation in two dimensions. For a spinor with polarization $u^\alpha$, $\alpha=\pm$, the equation reads
		\be \begin{cases}
			(k^0+k^1)u^+=0\\ (k^0-k^1)u^-=0
		\end{cases}
		\ee
		The states in $C(k)_{-1/2,1}$ have polarization $u^+=0$ and $u^-\neq 0$, so they are solutions of the Dirac equation only if $k^0=k^1$.
		\item for $k^0=k^1$, $H(k)_{-1/2,1}$ is $24$ dimensional, spanned by $(\chi_+\Theta_-e^{-\phi/2})_{-1}c_1|k\rangle$ (all states are in $\ker Q$ in this case). This is true in particular for $k=0$.
		\item for $k^0= k^1$, one has $H(k)_{-3/2,1}=24$.  This includes in particular the case $k=0$. Indeed, all states in $C(k)_{-3/2,1}$ are in $\ker Q$. Furthermore, the operator $Q$ vanishes on a state in $C_{-3/2,0}(k)$, when $k^0=k^1$, so there are no exact states in $C(k)_{-3/2,1}$.
		\item for  $k^2=0$, $k^0\neq k^1$, we have $H(k)_{-3/2,1}=0$, because all states of $C(k)_{-3/2,1}$ are $Q$-exact  when $k^0=-k^1\neq 0$. 
	\end{itemize}

To summarize, we have the following cases for $k^2=0$ for the picture number $-1/2$
\be
\begin{matrix}
	& \dim H(k)_{-1/2,1} & \dim H(k)_{-3/2,1}\\
	k^0=-k^1\neq 0	& 0 & 0\\
	k^0=k^1	& 24 & 24
\end{matrix}
\ee

\section{Proof of identity \eqref{infprod}}\label{a:prfusefulid}

In this appendix, we will give a proof of \eqref{infprod}. We have
\begin{align}
\sum_{m=1}^\infty p^m(T_mf)(x,y;\tau)&=\sum_{a,d=1}^\infty\frac{p^{ad}}{ad}\sum_{b=0}^{d-1} f(dx,ay-bx;\frac{a\tau+b}{d})\\
&=\sum_{a,d=1}^\infty\frac{p^{ad}}{ad}\sum_{b=0}^{d-1}\sum_{n\in\ZZ} q^{\frac{an}{Nd}}e^{2\pi i\frac{bn}{Nd}}c_{dx,ay-bx}(\frac{n}{N})\\
&=\sum_{a,d=1}^\infty\frac{p^{ad}}{ad}\sum_{b=0}^{d-1}\sum_{n\in \ZZ} q^{\frac{an}{Nd}}e^{2\pi i\frac{bn}{Nd}}\sum_{t,l=0}^{N-1}\frac{1}{N}e^{2\pi i\frac{t(a-l)}{N}}c_{dx,ly-bx}(\frac{n}{N}).
\end{align}
We now use the identity (see below for a proof, adapted from a more general argument in \cite{Persson:2013xpa}, Appendix D.1) 
\be\label{usefulid}  \frac{1}{d}\sum_{b=0}^{d-1}\sum_{n\in \ZZ} q^{\frac{an}{Nd}}e^{2\pi i\frac{bn}{Nd}}c_{dx,ly-bx}(\frac{n}{N})= \frac{1}{N}\sum_{b=0}^{N-1}\sum_{r\in \ZZ} q^{\frac{ar}{N}}e^{2\pi i\frac{br}{N}}c_{dx,ly-bx}(\frac{rd}{N})
\ee	
to get
\begin{align}
\sum_{m=1}^\infty p^m(T_mf)(x,y;\tau)&=\sum_{d=1}^\infty\frac{1}{N}\sum_{b=0}^{N-1}\sum_{r\in \ZZ} e^{2\pi i\frac{br}{N}}\sum_{t,l=0}^{N-1}\frac{1}{N}e^{-2\pi i\frac{tl}{N}}c_{dx,ly-bx}(\frac{rd}{N})\sum_{a=1}^\infty\frac{(e^{2\pi i\frac{t}{N}}p^{d}q^{\frac{r}{N}})^a}{a}\\
&=\sum_{d=1}^\infty\sum_{r\in\ZZ} \sum_{t=0}^{N-1} \log(1-e^{2\pi i\frac{t}{N}}p^{d}q^{\frac{r}{N}})^{\sum_{k=0}^{N-1}\delta^{(N)}(r-kx)\delta^{(N)}(t-ky)C_{dx,k}(\frac{rd}{N})}
\end{align}
where 
\be \delta^{(N)}(n)=\begin{cases}
	1 & \text{if }n\equiv0\mod N\\
	0 & \text{otherwise}
\end{cases}
\ee and
\be C_{x,k}(n):=\frac{1}{N}\sum_{y=0}^{N-1}e^{-2\pi i\frac{y k}{N}}c_{x,y}(n)
\ee are the Fourier coefficients of $F_{x,k}(\tau)$ defined in \eqref{Fdef}. The last equality follows from
\begin{align} &\frac{1}{N^2}\sum_{b,l=0}^{N-1} e^{2\pi i\frac{br-tl}{N}}c_{dx,ly-bx}(n)=\frac{1}{N^2}\sum_{b,l,k=0}^{N-1} e^{2\pi i\frac{br-tl}{N}}e^{2\pi i\frac{k(ly-bx)}{N}}C_{dx,k}(n)\\
&= \sum_{k=0}^{N-1} \delta^{(N)}(r-kx)\delta^{(N)}(t-ky)C_{dx,k}(n)
\end{align}

Finally, let us show that eq.\eqref{usefulid} holds.
This proof is a adaptation of the proof in appendix D.1 of \cite{Persson:2013xpa}. First notice that the condition
\be f(dx,ly-bx;\tau+1)=f(dx,ly-bx+dx;\tau)\ ,
\ee implies
\be\label{periodic} c_{dx,ly-bx}(n)e^{2\pi i n}=c_{dx,ly-bx+dx}(n)\ .
\ee Let us define
\be e:=\gcd(d,N)\qquad f:=\frac{d}{e}\ ,
\ee and note that
\be\label{coprime} \gcd(\frac{N}{e},f)=1\ .
\ee We have $f(dx,ly-bx;\tau+N/e)=f(dx,ly-bx+Nx\frac{d}{e};\tau)=f(dx,ly-bx;\tau)$, since $d/e$ is an integer and $Nxd/e\equiv 0\mod N$; therefore,  $c_{dx,ly-bx}(\frac{n}{N})=0$ unless $e|n$. Thus, we can define $h:=n/e$ and obtain
\be  \frac{1}{d}\sum_{b=0}^{d-1}\sum_{n\in \ZZ} q^{\frac{an}{Nd}}e^{2\pi i\frac{bn}{Nd}}c_{dx,ly-bx}(\frac{n}{N})= \frac{1}{d}\sum_{b=0}^{d-1}\sum_{h\in \ZZ} q^{\frac{ah}{Nf}}e^{2\pi i\frac{bh}{Nf}}c_{dx,ly-bx}(\frac{he}{N})\ .
\ee By \eqref{periodic}, the general term in the sum over $b$ is periodic under $b\mapsto b+d$.
Let us set $b:=se+b'$ and replace the sum over $b\in \ZZ/d\ZZ$ by a sum over $b'\in \ZZ/e\ZZ$ and $s\in \ZZ/f\ZZ$
\be  \sum_{h\in \ZZ} \frac{1}{e}\sum_{b'\in \ZZ/e\ZZ} \frac{1}{f}\sum_{s\in \ZZ/f\ZZ} q^{\frac{ah}{Nf}}e^{2\pi i\frac{(b'+se)h}{Nf}}c_{dx,ly-b'x-sex}(\frac{he}{N})\ .
\ee In this expression, the general term of the sum over $b'$ is invariant under $b'\mapsto b'+e$, so we can replace $\frac{1}{e}\sum_{b'\in\ZZ/e\ZZ}$ by $\frac{1}{N}\sum_{b'\in \ZZ/N\ZZ}$ to get
\be  \sum_{h\in \ZZ} \frac{1}{N}\sum_{b'\in \ZZ/N\ZZ} \frac{1}{f}\sum_{s\in \ZZ/f\ZZ} q^{\frac{ah}{Nf}}e^{2\pi i\frac{(b'+se)h}{Nf}}c_{dx,ly-b'x-sex}(\frac{he}{N})\ .
\ee
By \eqref{coprime}, there are integers $u,v$ such that
\be\label{copr} uf+v\frac{N}{e}=1\ ,
\ee so that $e=ud+vN$ and
\begin{align} &\sum_{h\in \ZZ} \sum_{b'\in \ZZ/N\ZZ} \frac{e^{2\pi i\frac{b'h}{Nf}}}{N}\frac{1}{f}\sum_{s\in \ZZ/f\ZZ} q^{\frac{ah}{Nf}}e^{2\pi i\frac{s(ud+vN)h}{Nf}}c_{dx,ly-b'x-sudx}(\frac{he}{N})\\
&=\sum_{h\in \ZZ} \sum_{b'\in \ZZ/N\ZZ} \frac{e^{2\pi i\frac{b'h}{Nf}}}{N}\frac{1}{f}\sum_{s\in \ZZ/f\ZZ} q^{\frac{ah}{Nf}}e^{2\pi i\frac{svh}{f}}c_{dx,ly-b'x}(\frac{he}{N})
\end{align} where we used that, by \eqref{periodic}, $e^{2\pi i\frac{sueh}{N}}c_{dx,ly-b'x-sudx}(\frac{he}{N})=c_{dx,ly-b'x}(\frac{he}{N})$. Since, by \eqref{copr}, $\gcd(v,f)=1$, we get
\be \frac{1}{f}\sum_{s\in \ZZ/f\ZZ} e^{2\pi i\frac{svh}{f}}=\begin{cases}
	1 & \text{if }h\equiv0\mod f\\ 
	0 & \text{otherwise.}
\end{cases}
\ee Therefore, the sum over $s$ is non-vanishing only if $f|h$ and we can set $h=rf$ and obtain
\be \frac{1}{d}\sum_{b=0}^{d-1}\sum_{n\in \ZZ} q^{\frac{an}{Nd}}e^{2\pi i\frac{bn}{Nd}}c_{dx,ly-bx}(\frac{n}{N})=\sum_{r\in \ZZ} \sum_{b'\in \ZZ/N\ZZ} \frac{e^{2\pi i\frac{b'r}{N}}}{N}q^{\frac{ar}{N}}c_{dx,ly-b'x}(\frac{rd}{N})\ ,
\ee which is equivalent to \eqref{usefulid}.


\begin{thebibliography}{9}

\bibitem{ACH} 
  V.~Anagiannis, M.~C.~N.~Cheng and S.~M.~Harrison,
  ``K3 Elliptic Genus and an Umbral Moonshine Module,''
  arXiv:1709.01952 [hep-th].
  
  
\bibitem{Distleretal} 
  A.~Bergman, J.~Distler and U.~Varadarajan,
  ``(1+1) dimensional critical string theory and holography,''
  hep-th/0312115. 

\bibitem{borcherds1988generalized}
  R.~E.~Borcherds
  ``Generalized Kac-Moody algebras,''
  Journal of Algebra {\bf 115}, 2 (1988)

\bibitem{BorcherdsFake}
  R.~E.~Borcherds,
``The monster {L}ie algebra,''
Adv.\ Math.\  {\bf 83}, 1
(1990).

\bibitem{BorcherdsMM}
  R.~E.~Borcherds,
  ``Monstrous moonshine and monstrous Lie superalgebras,''
  Invent.\ Math.\  {\bf 109}, 1 (1992)

\bibitem{borcherds1995automorphic}
  R.~E.~Borcherds,
  ``Automorphic forms on $O_{s+2, 2}(R)$ and infinite products,''
 Invent.\ Math.\  {\bf 120}, 1 (1995)
 


\bibitem{Borcherds:1996uda} 
  R.~E.~Borcherds,
  ``Automorphic forms with singularities on Grassmannians,''
  Invent.\ Math.\  {\bf 132}, 491 (1998)
  doi:10.1007/s002220050232
  [alg-geom/9609022].
  
\bibitem{Carnahan}
  S.~Carnahan,
  ``Generalized moonshine, II: Borcherds products,''
  Duke Mathematical Journal {\bf 161}, 5 (2012)
  
\bibitem{Carnahan2}
  S.~Carnahan,
  ``Generalized moonshine IV: monstrous Lie algebras,''
   arXiv:1208.6254
   
   \bibitem{Cheng:2010pq} 
  M.~C.~N.~Cheng,
  ``K3 Surfaces, N=4 Dyons, and the Mathieu Group M24,''
  Commun.\ Num.\ Theor.\ Phys.\  {\bf 4}, 623 (2010)
  doi:10.4310/CNTP.2010.v4.n4.a2
  [arXiv:1005.5415 [hep-th]].
   
 \bibitem{M5} 
  M.~C.~N.~Cheng, X.~Dong, J.~F.~R.~Duncan, S.~Harrison, S.~Kachru and T.~Wrase,
  ``Mock Modular Mathieu Moonshine Modules,''
  Res.\ Math.\ Sci.\  {\bf 2}, 13 (2015)
  doi:10.1186/s40687-015-0034-9
  [arXiv:1406.5502 [hep-th]].
   
\bibitem{CDRad} 
  M.~C.~N.~Cheng and J.~F.~R.~Duncan,
  ``On Rademacher Sums, the Largest Mathieu Group, and the Holographic Modularity of Moonshine,''
  Commun.\ Num.\ Theor.\ Phys.\  {\bf 6}, 697 (2012)
  doi:10.4310/CNTP.2012.v6.n3.a4
  [arXiv:1110.3859 [math.RT]].
   
\bibitem{CDRad2} 
  M.~C.~N.~Cheng and J.~F.~R.~Duncan,
  ``Rademacher Sums and Rademacher Series,''
  Contrib.\ Math.\ Comput.\ Sci.\  {\bf 8}, 143 (2014)
  doi:10.1007/978-3-662-43831-2-6
  [arXiv:1210.3066 [math.NT]].
  
   
\bibitem{CDOpt} 
  M.~C.~N.~Cheng and J.~F.~R.~Duncan,
  ``Optimal Mock Jacobi Theta Functions,''
  arXiv:1605.04480 [math.NT].
  
\bibitem{CDH1} 
  M.~C.~N.~Cheng, J.~F.~R.~Duncan and J.~A.~Harvey,
  ``Umbral Moonshine,''
  Commun.\ Num.\ Theor.\ Phys.\  {\bf 08}, 101 (2014)
  doi:10.4310/CNTP.2014.v8.n2.a1
  [arXiv:1204.2779 [math.RT]].
  
\bibitem{CDH2} 
  M.~C.~N.~Cheng, J.~F.~R.~Duncan and J.~A.~Harvey,
  ``Umbral Moonshine and the Niemeier Lattices,''
  arXiv:1307.5793 [math.RT].
  
\bibitem{CDMod} 
  M.~C.~N.~Cheng and J.~F.~R.~Duncan,
  ``Meromorphic Jacobi Forms of Half-Integral Index and Umbral Moonshine Modules,''
  arXiv:1707.01336 [math.RT].
   
\bibitem{CDHJacobi} 
  M.~C.~N.~Cheng, J.~F.~R.~Duncan and J.~A.~Harvey,
  ``Weight One Jacobi Forms and Umbral Moonshine,''
  J.\ Phys.\ A {\bf 51}, no. 10, 104002 (2018)
  doi:10.1088/1751-8121/aaa819
  [arXiv:1703.03968 [math.NT]].
  
\bibitem{CH} 
  M.~C.~N.~Cheng and S.~Harrison,
  ``Umbral Moonshine and K3 Surfaces,''
  Commun.\ Math.\ Phys.\  {\bf 339}, no. 1, 221 (2015)
  doi:10.1007/s00220-015-2398-5
  [arXiv:1406.0619 [hep-th]].
  
\bibitem{Equivariant} 
  M.~C.~N.~Cheng, J.~F.~R.~Duncan, S.~M.~Harrison and S.~Kachru,
  ``Equivariant K3 Invariants,''
  Commun.\ Num.\ Theor.\ Phys.\  {\bf 11}, 41 (2017)
  doi:10.4310/CNTP.2017.v11.n1.a2
  [arXiv:1508.02047 [hep-th]].
  
\bibitem{CHVZ} 
  M.~C.~N.~Cheng, S.~M.~Harrison, R.~Volpato and M.~Zimet,
  ``K3 String Theory, Lattices and Moonshine,''
  arXiv:1612.04404 [hep-th].
  
  \bibitem{CN}
J. H. Conway and S. P. Norton, ``Monstrous moonshine," Bull. London Math. Soc.
{\bf 11}, 308 (1979).
  
  \bibitem{CDR} 
  T.~Creutzig, J.~F.~R.~Duncan and W.~Riedler,
  ``Self-Dual Vertex Operator Superalgebras and Superconformal Field Theory,''
  J.\ Phys.\ A {\bf 51}, no. 3, 034001 (2018)
  doi:10.1088/1751-8121/aa9af5
  [arXiv:1704.03678 [math-ph]].
  
\bibitem{Gannon}
  C.J.~Cummins and T.~Gannon,
  ``Modular equations and the genus zero property of moonshine functions,''
  Inventiones mathematicae {\bf 129}, no. 3 (1997)
    
\bibitem{DJS} 
  J.~R.~David, D.~P.~Jatkar and A.~Sen,
  ``Dyon spectrum in generic N=4 supersymmetric Z(N) orbifolds,''
  JHEP {\bf 0701}, 016 (2007)
  doi:10.1088/1126-6708/2007/01/016
  [hep-th/0609109].

\bibitem{DVV} 
  R.~Dijkgraaf, E.~P.~Verlinde and H.~L.~Verlinde,
  ``Counting dyons in N=4 string theory,''
  Nucl.\ Phys.\ B {\bf 484}, 543 (1997)
  doi:10.1016/S0550-3213(96)00640-2
  [hep-th/9607026].
    

  
\bibitem{Duncan}
  J.~F.~Duncan
  ``Super-moonshine for Conway's largest sporadic group,''
  Duke Mathematical Journal {\bf 139}, 2 (2007)

\bibitem{DF} 
  J.~F.~Duncan and I.~B.~Frenkel,
  ``Rademacher sums, Moonshine and Gravity,''
  Commun.\ Num.\ Theor.\ Phys.\  {\bf 5}, 849 (2011)
  doi:10.4310/CNTP.2011.v5.n4.a4
  [arXiv:0907.4529 [math.RT]].
  
\bibitem{DH} 
  J.~F.~R.~Duncan and J.~A.~Harvey,
  ``The Umbral Moonshine Module for the Unique Unimodular Niemeier Root System,''
  Alg. Number Th. 11 (2017) 505-535
  doi:10.2140/ant.2017.11.505
  [arXiv:1412.8191 [math.RT]].

\bibitem{Duncan:2014eha} 
  J.~F.~R.~Duncan and S.~Mack-Crane,
  ``The Moonshine Module for Conway's Group,''
  SIGMA {\bf 3}, e10 (2015)
  doi:10.1017/fms.2015.7
  [arXiv:1409.3829 [math.RT]]. 
  
\bibitem{DuncanMC} 
  J.~F.~R.~Duncan and S.~Mack-Crane,
  ``Derived Equivalences of K3 Surfaces and Twined Elliptic Genera,''
  arXiv:1506.06198 [math.RT].
  
\bibitem{DOD} 
  J.~F.~R.~Duncan and A.~O'Desky,
  ``Super Vertex Algebras, Meromorphic Jacobi Forms and Umbral Moonshine,''
  arXiv:1705.09333 [math.RT].
  
\bibitem{EOT} 
  T.~Eguchi, H.~Ooguri and Y.~Tachikawa,
  ``Notes on the K3 Surface and the Mathieu group $M_{24}$,''
  Exper.\ Math.\  {\bf 20}, 91 (2011)
  doi:10.1080/10586458.2011.544585
  [arXiv:1004.0956 [hep-th]].
  
\bibitem{SarahFrancesca} 
  F.~Ferrari and S.~M.~Harrison,
  ``Properties of extremal CFTs with small central charge,''
  arXiv:1710.10563 [hep-th].

\bibitem{FLM0}
I. Frenkel, J. Lepowsky and A. Meurman, ``A moonshine module for the {M}onster.'' In
 {\it Vertex operators in mathematics and physics} ({B}erkeley,
{C}alif., 1983),
Math. Sci. Res. Inst. Publ.
{\bf 3}, Springer, New York, 1985.

  \bibitem{FLM}
I. Frenkel, J. Lepowsky and A. Meurman, {\it Vertex~ Operator ~Algebras ~and ~the ~Monster},
vol. 134 of Pure and Applied Mathematics, Elsevier Science, 1989.
  
\bibitem{GHV} 
  M.~R.~Gaberdiel, S.~Hohenegger and R.~Volpato,
  ``Symmetries of K3 sigma models,''
  Commun.\ Num.\ Theor.\ Phys.\  {\bf 6}, 1 (2012)
  doi:10.4310/CNTP.2012.v6.n1.a1
  [arXiv:1106.4315 [hep-th]].
  
  \bibitem{GarlandLepowsky}
  H.~Garland and J.~Lepowsky,
  ``Lie algebra homology and the Macdonald-Kac formulas,''
  Inventiones mathematicae {\bf 34}, 1 (1976).
  
 \bibitem{Goddard:1972iy} 
 P.~Goddard and C.~B.~Thorn,
 ``Compatibility of the Dual Pomeron with Unitarity and the Absence of Ghosts in the Dual Resonance Model,''
 Phys.\ Lett.\  {\bf 40B}, 235 (1972).
 
\bibitem{GN1} 
  V.~A.~Gritsenko and V.~V.~Nikulin,
  ``Automorphic forms and Lorentzian Kac-Moody algebras. Part 1.,''
  International Journal of Mathematics {\bf 9}, 2 (1998).
  
\bibitem{GN2} 
  V.~A.~Gritsenko and V.~V.~Nikulin,
  ``Automorphic forms and Lorentzian Kac-Moody algebras. Part 2,''
  International Journal of Mathematics {\bf 9}, 2 (1998).
  
  \bibitem{phi12} 
  S.~M.~Harrison, S.~Kachru, N.~M.~Paquette, R.~Volpato and M.~Zimet,
  ``Heterotic sigma models on $T^8$ and the Borcherds automorphic form $\Phi_{12}$,''
  JHEP {\bf 1710}, 121 (2017)
  doi:10.1007/JHEP10(2017)121
  [arXiv:1610.00707 [hep-th]].
  
\bibitem{HM1} 
  J.~A.~Harvey and G.~W.~Moore,
  ``Algebras, BPS states, and strings,''
  Nucl.\ Phys.\ B {\bf 463}, 315 (1996)
  doi:10.1016/0550-3213(95)00605-2
  [hep-th/9510182].
  
\bibitem{HM2} 
  J.~A.~Harvey and G.~W.~Moore,
  ``On the algebras of BPS states,''
  Commun.\ Math.\ Phys.\  {\bf 197}, 489 (1998)
  doi:10.1007/s002200050461
  [hep-th/9609017].
  
\bibitem{HM3} 
  J.~A.~Harvey and G.~W.~Moore,
  ``Conway Subgroup Symmetric Compactifications of Heterotic String,''
  arXiv:1712.07986 [hep-th].
  
\bibitem{jurisich1996exposition}
  E.~Jurisich,
  ``An exposition of generalized Kac-Moody algebras,''
  Contemporary Mathematics {\bf 194} (1996)
  
  \bibitem{jurisich1998generalized}
  E.~Jurisich,
  ``Generalized Kac-Moody Lie algebras, free Lie algebras and the structure of the Monster Lie algebra,''
  Journal of Pure and Applied Algebra {\bf 126} 1-3 (1998)

\bibitem{KPV} 
  S.~Kachru, N.~M.~Paquette and R.~Volpato,
  ``3D String Theory and Umbral Moonshine,''
  J.\ Phys.\ A {\bf 50}, no. 40, 404003 (2017)
  doi:10.1088/1751-8121/aa6e07
  [arXiv:1603.07330 [hep-th]].
  
\bibitem{KP} 
  T.~Kimura and V.~Pestun,
  ``Quiver W-algebras,''
  arXiv:1512.08533 [hep-th].
  
\bibitem{KS1} 
  M.~Kontsevich and Y.~Soibelman,
  ``Stability structures, motivic Donaldson-Thomas invariants and cluster transformations,''
  arXiv:0811.2435 [math.AG].
  
\bibitem{KS2} 
  M.~Kontsevich and Y.~Soibelman,
  ``Cohomological Hall algebra, exponential Hodge structures and motivic Donaldson-Thomas invariants,''
  Commun.\ Num.\ Theor.\ Phys.\  {\bf 5}, 231 (2011)
  doi:10.4310/CNTP.2011.v5.n2.a1
  [arXiv:1006.2706 [math.AG]].
  




  
\bibitem{LZ}
B.~H.~Lian and G.~J.~Zuckerman,
``{BRST} Cohomology of the Supervirasoro Algebras,''
Commun.\ Math.\ Phys.\  {\bf 125} (1989) 301.
doi:10.1007/BF01217910


  
  \bibitem{LZ2}
  B.~H.~Lian and G.~J.~Zuckerman,
  ``New perspectives on the BRST algebraic structure of string theory,''
  Commun.\ Math.\ Phys.\  {\bf 154} (1993) 613
  doi:10.1007/BF02102111
  [hep-th/9211072].
  
\bibitem{PPV} 
  N.~M.~Paquette, D.~Persson and R.~Volpato,
  ``Monstrous BPS-Algebras and the Superstring Origin of Moonshine,''
  Commun.\ Num.\ Theor.\ Phys.\  {\bf 10}, 433 (2016)
  doi:10.4310/CNTP.2016.v10.n3.a2
  [arXiv:1601.05412 [hep-th]].
  
\bibitem{PPV2} 
  N.~M.~Paquette, D.~Persson and R.~Volpato,
  ``BPS Algebras, Genus Zero, and the Heterotic Monster,''
  J.\ Phys.\ A {\bf 50}, no. 41, 414001 (2017)
  doi:10.1088/1751-8121/aa8443
  [arXiv:1701.05169 [hep-th]].
  
\bibitem{PVZ} 
  N.~M.~Paquette, R.~Volpato and M.~Zimet,
  ``No More Walls! A Tale of Modularity, Symmetry, and Wall Crossing for 1/4 BPS Dyons,''
  JHEP {\bf 1705}, 047 (2017)
  doi:10.1007/JHEP05(2017)047
  [arXiv:1702.05095 [hep-th]].
  
  \bibitem{Persson:2015jka} 
  D.~Persson and R.~Volpato,
  ``Fricke S-duality in CHL models,''
  JHEP {\bf 1512}, 156 (2015)
  doi:10.1007/JHEP12(2015)156
  [arXiv:1504.07260 [hep-th]].
  
\bibitem{Persson:2013xpa}
D.~Persson and R.~Volpato,
``Second Quantized Mathieu Moonshine,''
Commun.\ Num.\ Theor.\ Phys.\  {\bf 08} (2014) 403
doi:10.4310/CNTP.2014.v8.n3.a2
[arXiv:1312.0622 [hep-th]].
 
 \bibitem{Polchinski:1998rr}
 J.~Polchinski,
 ``String theory. Vol. 2: Superstring theory and beyond,'' Cambridge University Press (1998).

\bibitem{Ray} 
  U.~Ray,
  ``Automorphic Forms and Lie Superalgebras,''
   Springer Science \& Business Media (2007).
  
\bibitem{Sch1} 
N.~R.~Scheithauer,
``The Fake monster superalgebra,''
Adv. Math. {\bf  151}, no. 2 (2000) {\tt [arXiv:math/9905113]}.

\bibitem{Sch2}  N.~R.~Scheithauer,
  ``Vertex algebras, Lie algebras and superstrings,''
  Journal of Algebra {\bf 200}, no. 2 (1998)
  [arXiv:9802058 [hep-th]].
  
  \bibitem{Sch5}  N.~R.~Scheithauer,
``Twisting the fake {M}onster superalgebra,''
Adv. Math. {\bf  164}, no. 2 (2001).
  
\bibitem{Sch3}  N.~R.~Scheithauer,
``Generalized {K}ac-{M}oody algebras, automorphic forms and
{C}onway's group. {I}'',
Adv.\ Math.\  {\bf 183}, 2
(2004). 

\bibitem{Sch4}  N.~R.~Scheithauer,
``Generalized {K}ac-{M}oody algebras, automorphic forms and
{C}onway's group. {II}'',
J.\ Reine\ Angew.\ Math. {\bf 625},
(2008).

\bibitem{Taormina:2017zlm} 
A.~Taormina and K.~Wendland,
``The Conway Moonshine Module is a Reflected K3 Theory,''
arXiv:1704.03813 [hep-th].

\bibitem{Thompson1}
J.~G.~Thompson, 
``Finite groups and modular functions,''
Bull. London Math. Soc.
{\bf 11}, 347 (1979).

\bibitem{Thompson2}
J.~G.~Thompson, 
``Some numerology between the {F}ischer-{G}riess {M}onster and
the elliptic modular function,''
Bull. London Math. Soc.
{\bf 11}, 352 (1979).


\end{thebibliography}
\end{document}